\documentclass[aps,pre,twocolumn,10pt,superscriptaddress,showpacs]{revtex4-1}

\def\bra#1{\mathinner{\langle{#1}|}}
\def\ket#1{\mathinner{|{#1}\rangle}}
\def\braket#1{\mathinner{\langle{#1}\rangle}}

\let\protect\relax
{\catcode`\|=\active
  \xdef\Braket{\protect\expandafter\noexpand\csname Braket \endcsname}
  \expandafter\gdef\csname Braket \endcsname#1{\begingroup
     \ifx\SavedDoubleVert\relax
       \let\SavedDoubleVert\|\let\|\BraDoubleVert
     \fi
     \mathcode`\|32768\let|\BraVert
     \left\langle{#1}\right\rangle\endgroup}
}
\def\BraVert{\@ifnextchar|{\|\@gobble}
     {\egroup\,\mid@vertical\,\bgroup}}
\def\BraDoubleVert{\egroup\,\mid@dblvertical\,\bgroup}
\let\SavedDoubleVert\relax

{\catcode`\|=\active
  \xdef\set{\protect\expandafter\noexpand\csname set \endcsname}
  \expandafter\gdef\csname set \endcsname#1{\mathinner
        {\lbrace\,{\mathcode`\|32768\let|\midvert #1}\,\rbrace}}
  \xdef\Set{\protect\expandafter\noexpand\csname Set \endcsname}
  \expandafter\gdef\csname Set \endcsname#1{\left\{%
     \ifx\SavedDoubleVert\relax \let\SavedDoubleVert\|\fi
     \:{\let\|\SetDoubleVert
     \mathcode`\|32768\let|\SetVert
     #1}\:\right\}}
}
\def\midvert{\egroup\mid\bgroup}
\def\SetVert{\@ifnextchar|{\|\@gobble}
    {\egroup\;\mid@vertical\;\bgroup}}
\def\SetDoubleVert{\egroup\;\mid@dblvertical\;\bgroup}

\begingroup
 \edef\@tempa{\meaning\middle}
 \edef\@tempb{\string\middle}
\expandafter \endgroup \ifx\@tempa\@tempb
 \def\mid@vertical{\middle|}
 \def\mid@dblvertical{\middle\SavedDoubleVert}
\else
 \def\mid@vertical{\mskip1mu\vrule\mskip1mu}
 \def\mid@dblvertical{\mskip1mu\vrule\mskip2.5mu\vrule\mskip1mu}
\fi

\usepackage{graphicx}
\usepackage[T1]{fontenc}
\usepackage[]{amsfonts}
\usepackage[]{amsmath}
\usepackage[]{rotating}
\usepackage[]{color}
\usepackage[]{float}
\usepackage[]{fancyhdr}
\usepackage[]{booktabs}
\usepackage[]{epsfig}
\usepackage{appendix}
\usepackage{amssymb}
\usepackage{psfrag}
\usepackage{setspace}

\begin{document}

\title{Fluctuations of work in nearly adiabatically driven open quantum systems}

\author{S. Suomela}
\affiliation{COMP Centre of Excellence, Department of Applied Physics, Aalto University, P.O. Box 11000, FI-00076 Aalto, Finland}

\author{J. Salmilehto}
\affiliation{QCD Labs, COMP Centre of Excellence, Department of Applied Physics, Aalto University, P.O. Box 13500, FI-00076 Aalto, Finland}
\affiliation{Department of Physical Chemistry, University of the Basque Country UPV/EHU, Apartado 644, 48080 Bilbao, Spain}

\author{I. G. Savenko}
\affiliation{COMP Centre of Excellence, Department of Applied Physics, Aalto University, P.O. Box 11000, FI-00076 Aalto, Finland}
\affiliation{QCD Labs, COMP Centre of Excellence, Department of Applied Physics, Aalto University, P.O. Box 13500, FI-00076 Aalto, Finland}
\affiliation{Low Temperature Laboratory (OVLL), Aalto University, P.O. Box 13500, FI-00076 Aalto, Finland}
\affiliation{National Research University of Information Technologies, Mechanics and Optics (ITMO University), Saint-Petersburg 197101, Russia}

\author{T. Ala-Nissila}
\affiliation{COMP Centre of Excellence, Department of Applied Physics, Aalto University, P.O. Box 11000, FI-00076 Aalto, Finland}
\affiliation{Department of Physics, P.O. Box 1843, Brown University, Providence, Rhode Island 02912-1843, USA}

\author{M. M\"ott\"onen}
\affiliation{QCD Labs, COMP Centre of Excellence, Department of Applied Physics, Aalto University, P.O. Box 13500, FI-00076 Aalto, Finland}
\affiliation{Low Temperature Laboratory (OVLL), Aalto University, P.O. Box 13500, FI-00076 Aalto, Finland}

\date{\today}

\begin{abstract}

We extend the quantum jump method to nearly adiabatically driven open quantum systems in a way that allows for an accurate account of the external driving in the system--environment interaction. Using this framework, we construct the corresponding trajectory-dependent work performed on the system and derive the integral fluctuation theorem and the Jarzynski equality for nearly adiabatic driving. We show that such identities hold as long as the stochastic dynamics and work variable are consistently defined. We numerically study the emerging work statistics for a two-level quantum system and find that the conventional diabatic approximation is unable to capture some prominent features arising from driving such as the continuity of the probability density of work. Our results reveal the necessity of using accurate expressions for the drive-dressed heat exchange in future experiments probing jump time distributions.

\end{abstract}


\maketitle

\section{Introduction}

The last two decades have witnessed great advancements in nonequilibrium thermodynamics and near-adiabatic evolution. In particular, fluctuation relations ~\cite{spjetp45/125,prl78/2690,Seifert2005} have opened a new way to tackle the obscure nature of nonequilibrium processes. Perhaps the most well-known of these is the Jarzynski equality~\cite{prl78/2690} that connects the average exponentiated work performed on a driven system with the free-energy difference between the initial and final thermal distributions.

In recent years, there have been numerous theoretical approaches aimed toward extending the classical fluctuation relations to the quantum regime~\cite{arxiv/0007360,arxiv/0009244,jpsj69/2367,prl90/170604,prl93/048302,pre71/066102,pre75/050102, prl100/230404,Engel2007,Schroder2007,epl83/30008,PhysRevE.77.051131,njp15/115008,PhysRevE.90.022107,PhysRevE.90.032137} with increasing emphasis on open quantum systems~\cite{pre73/046129, prl102/210401, PhysRevLett.107.140404,Crooks2007, jsp148/480, pre88/032146, rmp81/1665,rmp83/771, jsm/P06016, prb87/060508, prl111/093602, pre89/012127, pre89/032114, PhysRevE.89.052119, prb90/094304, PhysRevE.90.022103, PhysRevE.89.042122, PhysRevB.90.075421, 1367-2630-16-11-115001, PhysRevE.85.031110,NJP15/085028,PhysRevA.88.042111,PhysRevE.89.052128}. Typically the definition of work includes the change in the internal energy of the system obtained using the so-called two-measurements approach (TMA)~\cite{arxiv/0007360,arxiv/0009244} based on performing von Neumann measurements of the system energy at the beginning and at the end of the protocol. Due to difficulties in reliably implementing these measurements to date, the only experimental realizations of work distributions in essentially closed quantum systems~\cite{prl113/140601} have been achieved with an alternative method based on the characteristic function of the work distribution~\cite{prl110/230601,prl110/230602}. However, promising approaches to overcome these difficulties have been suggested~\cite{prl101/070403,prl108/190601,njp15/115006}.

For open quantum systems, the heat exchange between the system and a heat bath is added to the corresponding internal energy change to obtain the work performed. To model the work statistics, the quantum jump method can be used, as previously studied for the Lindblad equation with a time-independent dissipative part~\cite{prl111/093602,prb90/094304, PhysRevE.90.022103, PhysRevE.89.042122}. However, in the presence of a strong drive, the dissipative part may become time-dependent, which has been proven to affect the fluctuation relations~\cite{jsp148/480,PhysRevE.85.031110,NJP15/085028,PhysRevA.88.042111}. In addition, the heat value associated with the transitions may change.

The effect of time-dependent dissipation on work has been previously studied theoretically in the adiabatic limit~\cite{prl90/170604,PhysRevE.90.032121}. However, these studies do not provide quantitative estimates on the correction induced by the finite speed of driving on the work distributions. To advance beyond the bounds of the adiabatic limit, the adiabatic renormalization~\cite{prsla414/31, GPIPberry} provides an effective tool to study the dynamics in the case of nearly adiabatic driving. The adiabatic renormalization procedure can be used for open quantum systems by accurately taking into account the effect of driving on the dissipation in the derivation of the master equation~\cite{prb82/134517,pra82/062112,prb84/174507}. In the adiabatic renormalization procedure for open quantum systems, the error related to the time-local adiabatic parameter typically decreases quickly with increasing number of basis rotations~\cite{prb84/174507}. Thus for nearly adiabatic driving, a rather low number of basis rotations is sufficient.

In this paper, we extend the quantum jump method to nearly adiabatically driven open quantum systems by uniting the quantum jump theory with a master equation formalism utilizing the adiabatic renormalization. Thus we may consistently account for the influence of external driving on the system--environment interaction. Using this framework, we study the nonequilibrium work relations for nearly adiabatic driving. To this end, we construct the corresponding trajectory-dependent work and derive the integral fluctuation theorem and the Jarzynski equality. We show that such identities hold as long as the stochastic dynamics and work variable are consistently defined. To illustrate the results, we  consider a sinusoidally driven open two-level system with a large driving amplitude. We investigate the resulting work statistics for the zeroth, first, and second order adiabatic renormalization. We observe that the conventional diabatic approximation for the dissipative transitions is unable to capture some of the prominent features arising from driving, such as the continuity of the probability density of work. 

Our theoretical predictions are potentially experimentally observable, for example, with the calorimetric scheme~\cite{njp15/115006} providing the amount of heat released from the system to its environment. However, our results are even more important for schemes, in which only the jump times are measured~\cite{prl109/180601, np9/644, prl113/030601} together with the system Hamiltonian. In this case, the experimental dynamics is accurately described by a great number of basis rotations in the adiabatic renormalization scheme, but if a high-order theory presented in this paper is not employed to extract the work distributions, the resulting fluctuation relations are not precisely satisfied.

The paper is organized as follows. In Sec. \ref{sec:AR}, we provide a short summary of the adiabatic renormalization procedure. In Sec. \ref{sec:DDD}, we present the master equation for the driven system utilizing the renormalization. In Sec. \ref{sec:2lvl}, we introduce the two-level model system and analytically derive the quantities relevant to the master equation. In Sec. \ref{sec:Qtraj}, we apply the quantum jump method to the master equation of Sec. \ref{sec:DDD} and derive the resulting fluctuation relations. In Sec. \ref{sec:num}, we present the numerical analysis of the work statistics. The conclusions are drawn in Sec. \ref{sec:con}.

\section{Adiabatic renormalization} \label{sec:AR}

We examine a quantum system described by the Hamiltonian $\hat{H}_S(t)$, where the time dependence is caused by the action of external control fields inducing driven dynamics. In addition, the system is coupled to an environment such that the total Hamiltonian reads
\begin{equation}
\hat{H}(t) = \hat{H}_S(t) + \hat{V} + \hat{H}_E,
\label{eq:H}
\end{equation}
where $\hat{V}$ is the system-environment coupling operator and $\hat{H}_E$ is the environment Hamiltonian. We further assume that the coupling operator factorizes as $\hat{V} = \hat{Y} \otimes \hat{X}$, where $\hat{Y}$ and $\hat{X}$ operate in the system and the environment Hilbert spaces, respectively.

To access the dissipative dynamics for nearly adiabatic driving, we apply the adiabatic renormalization procedure~\cite{prsla414/31, GPIPberry} as introduced in Ref.~\cite{prb84/174507} in the context of open quantum systems. This allows for a basis selection that can be used to consistently account for the joint effect of driving and dissipation. Let us diagonalize the Hamiltonian $\hat{H}_S(t)$ in a time-independent basis $\{\ket{m}\}$ using the eigendecomposition as $\hat{\tilde{H}}^{(1)}_S(t)=\hat{D}_1^{\dagger}(t)\hat{H}_S(t)\hat{D}_1(t)$ corresponding to the eigenproblem $\hat{H}_S(t) \ket{m^{(1)}(t)} = E_m^{(1)}(t)\ket{m^{(1)}(t)}$, where $\hat{D}_1 \ket{m} = \ket{m^{(1)}(t)}$ is normalized and nondegenerate for each $m$. The states $\{ \ket{m} \}$ are here referred to as diabatic. If we similarly transform the total density operator $\hat{\rho}(t)$ in the Schr\"odinger picture as $\hat{\tilde{\rho}}^{(1)}(t)=\hat{D}_1^{\dagger}(t)\hat{\rho}(t)\hat{D}_1(t)$, the evolution of $\hat{\tilde{\rho}}^{(1)}(t)$ is governed by the effective Hamiltonian~\cite{pra82/062112, prb84/174507}
\begin{equation}
\hat{\tilde{H}}^{(1)}(t) = \hat{\tilde{H}}^{(1)}_S(t) + \hbar \hat{w}_1(t) + \hat{\tilde{V}}^{(1)}(t) + \hat{H}_E,
\label{eq:Heff}
\end{equation}
where $\hat{\tilde{V}}^{(1)}(t)=\hat{D}_1^{\dagger}(t)\hat{V}(t)\hat{D}_1(t)=\hat{D}_1^{\dagger}(t)\hat{Y}\hat{D}_1(t) \otimes \hat{X}(t)$ and $\hat{w}_1(t) = -i\hat{D}_1^{\dagger}(t)\dot{\hat{D}}_1(t)$. The notation $\dot{a}$ denotes the explicit time-derivative of arbitrary quantity $a$.

If we define a further unitary transformation $\hat{D}_2(t)$ rendering $\hat{\tilde{H}}^{(1)}_S(t) + \hbar \hat{w}_1(t)$ diagonal in the diabatic basis, the evolution of the density matrix $\hat{\tilde{\rho}}^{(2)} = \hat{D}_2^{\dagger}\hat{\tilde{\rho}}^{(1)}\hat{D}_2 = \hat{D}_2^{\dagger}\hat{D}_1^{\dagger}\hat{\rho}\hat{D}_1\hat{D}_2$ is governed by~\cite{pra82/062112}
\begin{equation}
\hat{\tilde{H}}^{(2)}(t) = \hat{\tilde{H}}^{(2)}_S(t) + \hbar \hat{w}_2(t) + \hat{\tilde{V}}^{(2)}(t) + \hat{H}_E,
\label{eq:Heff2}
\end{equation}
where $\hat{\tilde{H}}^{(2)}_S(t)=\hat{D}_2^{\dagger}(t)[\hat{\tilde{H}}^{(1)}_S(t)+\hbar \hat{w}_1(t)]\hat{D}_2(t)$, $\hat{\tilde{V}}^{(2)}(t)=\hat{D}_2^{\dagger}(t)\hat{\tilde{V}}^{(1)}(t)\hat{D}_2(t)$, and $\hat{w}_2=-i\hat{D}_2^{\dagger}(t)\dot{\hat{D}}_2(t)$. After $n$ successive coordinate transformations defined in a similar manner, the corresponding density operator is expressed as $\hat{\tilde{\rho}}^{(n)} = (\hat{D}_S^{(n)})^{\dagger}\hat{\rho}\hat{D}_S^{(n)}$ and $\hat{D}_S^{(n)}=\prod_{i=1}^{n}\hat{D}_{i} = \hat{D}_1\hat{D}_2 \cdots \hat{D}_{n-1}\hat{D}_{n}$. The evolution of $\hat{\tilde{\rho}}^{(n)}$ is subsequently governed by an effective Hamiltonian of
\begin{equation}
\hat{\tilde{H}}^{(n)} = \hat{\tilde{H}}_S^{(n)} + \hbar \hat{w}_{n} + \hat{\tilde{V}}^{(n)} + \hat{H}_E,
\label{eq:nH}
\end{equation}
where $\hat{\tilde{H}}_S^{(n)} = \hat{D}_{n}^{\dagger}[\hat{\tilde{H}}_S^{(n-1)} + \hbar \hat{w}_{n-1}]\hat{D}_{n}$, $\hat{\tilde{V}}^{(n)} = (\hat{D}_S^{(n)})^{\dagger}\hat{V}\hat{D}_S^{(n)}$ and $\hat{w}_{n} = -i\hat{D}_{n}^{\dagger}\dot{\hat{D}}_{n}$. Note that each coordinate transformation defines a set of time-dependent basis states $\{\ket{m^{(n)}} = \hat{D}_S^{(n)}\ket{m}\}$ that iteratively provide a better approximation of the closed-system evolution for smooth nearly adiabatic driving in the sense that driving-induced transitions between the states are suppressed. For $n=1$, these states correspond to the eigenstates of $\hat{H}_S$, referred to as the adiabatic states. For $n\geq2$, the states correspond to the eigenstates of $\hat{D}_S^{(n-1)}[\hat{\tilde{H}}_S^{(n-1)}+\hbar\hat{w}_{n-1}](\hat{D}_S^{(n-1)})^{\dagger}$, referred to as the $(n-1)$th-order superadiabatic states.

\section{Driven dissipative dynamics} \label{sec:DDD}

We assume that the system--environment coupling in Eq.~(\ref{eq:H}) is weak, so that we can adopt the approach used to derive the master equations for nearly adiabatically driven two-level quantum systems introduced in Refs.~\cite{prl105/030401, prb82/134517, pra82/062112, prb84/174507} for the reduced system density operator $\hat{\rho}_S = \mathrm{Tr}_E \left\{ \hat{\rho} \right\}$. For a general two-level quantum system, we denote the diabatic states as $\ket{g}$ and $\ket{e}$. We employ the secular approximation~\cite{API, pra85/032110} in the derivation detailed in Appendix~\ref{app:der} and obtain
\begin{equation}
\begin{split}
\dot{\hat{\rho}}_S &= -\frac{i}{\hbar} \left[ \hat{H}_S,  \hat{\rho}_S \right] \\
 &+  \sum_{i=0}^2 \left( \hat{L}_{(n,i)} \hat{\rho}_S  \hat{L}_{(n,i)}^\dagger - \frac{1}{2}  \left\lbrace \hat{L}_{(n,i)}^\dagger \hat{L}_{(n,i)}   ,\hat{\rho}_S \right\rbrace \right),
\end{split}
\label{eq:ME}
\end{equation}
where the error is of the third order in the coupling strength and linear order in the local (super)adiabatic parameter $\alpha_n = ||\hat{w}_n||/\omega_{01}^{(n)}$, where $\hbar \omega_{01}^{(n)} = E_e^{(n)} - E_g^{(n)}$ such that $E_e^{(n)} = \braket{e|\hat{\tilde{H}}_S^{(n)}|e}$ and $E_g^{(n)} = \braket{g|\hat{\tilde{H}}_S^{(n)}|g}$. We denote the Hilbert-Schmidt norm by $||\hat{w}_n(t)|| = \sqrt{\textrm{Tr}_S\{\hat{w}_n(t)^{\dagger}\hat{w}_n(t)\}}$. In order for Eq. \eqref{eq:ME} to give an accurate approximation of the real dynamics, the local (super)adiabatic parameter must satisfy  $\alpha_n \ll 1$. For nearly adiabatic driving, this condition is usually satisfied already with  a rather low number of basis rotations. The Lindblad operators are given by
\begin{equation}
\begin{split}
\hat{L}_{(n,0)} &= \sqrt{\Gamma_{(n,0)}} \ket{g^{(n)}} \bra{e^{(n)}}, \\
\hat{L}_{(n,1)} &= \sqrt{\Gamma_{(n,1)}} \ket{e^{(n)}} \bra{g^{(n)}}, \\
\hat{L}_{(n,2)} &= \sqrt{\Gamma_{(n,2)}} (\ket{e^{(n)}} \bra{e^{(n)}}-\ket{g^{(n)}} \bra{g^{(n)}}),
\label{eq:L}
\end{split}
\end{equation}
where the transition rates take the form
\begin{equation}
\begin{split}
\Gamma_{(n,0)} &= S(\omega_{01}^{(n)}) |m_2^{(n)}|^2,  \\
\Gamma_{(n,1)} &= S(-\omega_{01}^{(n)}) |m_2^{(n)}|^2,  \\
\Gamma_{(n,2)} &= S(0) |m_1^{(n)}|^2,
\end{split}
\end{equation}
where $m_1^{(n)} = \braket{g^{(n)}|\hat{Y}|g^{(n)}} = -\braket{e^{(n)}|\hat{Y}|e^{(n)}}$, $m_2^{(n)} = \braket{g^{(n)}|\hat{Y}|e^{(n)}}$, and the reduced spectral density of the noise source is defined as $S(\omega) = \int_{-\infty}^{\infty} d\tau \mathrm{Tr}_E \{\hat{\rho}_E \hat{X}_I(\tau) \hat{X}_I(0)\} e^{i\omega \tau}/\hbar^2$. Here, we use the conventional notation for operators in the interaction picture with respect to $\hat{H}_E$ such that $\hat{X}_I(t) = e^{i\hat{H}_E t/\hbar} \hat{X} e^{-i\hat{H}_E t/\hbar}$. The system part of the noise coupling operator, $\hat{Y}$, is traceless in the two-state basis following the convention used in the master equation of Refs. ~\cite{pra82/062112, prb84/174507} without additional transformation steps. Above we assume that the system time scales are much longer than the environment autocorrelation time such that the Markov and adiabatic-rate approximations apply~\cite{prl105/030401, prb82/134517, pra82/062112, prb84/174507}.
\begin{figure}
\centering
\includegraphics[width=8.7cm]{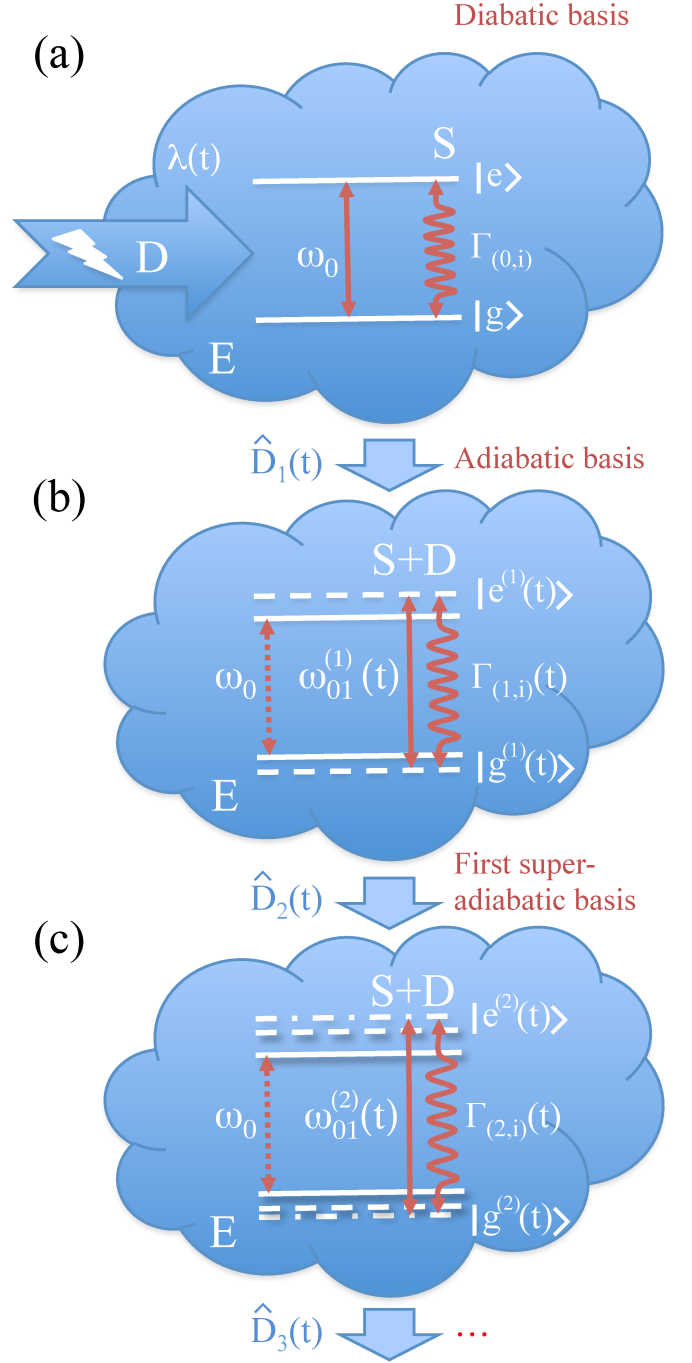}
\caption{(Color online) Schematic illustration of the effective energy level diagram of an externally driven two-level quantum system coupled to an environment. (a) Two-level system $S$ with level separation $\hbar \omega_0$ between $\ket{e}$ and $\ket{g}$ undergoing environment-induced transitions at rates $\Gamma_{(0,i)}$ experiences an external drive $D$. (b) After transformation $\hat{D}_1$, the environment induces transitions at rates $\Gamma_{(1,i)}$ between states $\ket{e^{(1)}}$ and $\ket{g^{(1)}}$ of the driven system $S+D$ separated by $\hbar \omega_{01}^{(1)}$. (c) Further transformation $\hat{D}_2$ yields transitions at rates $\Gamma_{(2,i)}$ between states $\ket{e^{(2)}}$ and $\ket{g^{(2)}}$ separated by $\hbar \omega_{01}^{(2)}$.}
\label{fig:model}
\end{figure}

Based on Appendix~\ref{app:der}, the unitary part of the Liouville operator in Eq.~(\ref{eq:ME}), which corresponds to the closed-system evolution, can be written as
\begin{equation}
\begin{split}
\hat{H}_S = & E_e^{(n)} \ket{e^{(n)}}\bra{e^{(n)}} + E_g^{(n)}\ket{g^{(n)}}\bra{g^{(n)}} \\ &+ \hbar w_{ge}^{(n)}(\ket{g^{(n)}}\bra{e^{(n)}}-\ket{e^{(n)}}\bra{g^{(n)}}) + \hbar \hat{W}_n,
\end{split}
\end{equation}
where $w_{kl}^{(n)} = \braket{k|\hat{w}_n|l}$, $\hat{W}_n = -i(\hat{D}_S^{(n)}) \partial_t [(\hat{D}_S^{(n)})^{\dagger}]$, and  we assumed $w_{gg}^{(n)} = w_{ee}^{(n)} = 0$ and $w_{eg}^{(n)} = (w_{ge}^{(n)})^* = -w_{ge}^{(n)}$ corresponding to the phase selection used in this paper~\cite{pra82/062112}. Furthermore, the total rotational term has the convenient identity in the $n$th-order basis representation of $\braket{k^{(n)}|\hat{W}_n|l^{(n)}} = -i\braket{\dot{k}^{(n)}|l^{(n)}}$. Note that if $\hat{w}_n$ is neglected due to slow driving, $\hat{H}_S \approx \hat{D}_S^{(n)}\hat{\tilde{H}}^{(n)}_S(\hat{D}_S^{(n)})^{\dagger} + \hbar \hat{W}_n$ yields (super)adiabatic evolution in the $n$th-order basis, that is, the driving-induced corrections to the closed-system state are fully accounted for by the basis selection and the corresponding transitions between the basis states are completely suppressed.

\section{Two-level model system} \label{sec:2lvl}

Our two-level model system is described by the Hamiltonian
\begin{equation}
\hat{H}_S(t) = \frac{\hbar\omega_0}{2} \hat{\sigma}_z+\lambda(t)(\hat{\sigma}_+ + \hat{\sigma}_-),
\end{equation}
where the Pauli operators are defined as $\hat{\sigma}_z = \ket{e}\bra{e} - \ket{g}\bra{g}$, $\hat{\sigma}_+ = \ket{e}\bra{g}$, $\hat{\sigma}_- = \ket{g}\bra{e}$, and $\omega_0$ is the resonance angular frequency. The states $\ket{g}$ and $\ket{e}$ are the ground and excited states of the undriven system ($\lambda = 0$) and we have conveniently selected them as the diabatic states. The system is steered in such a way that the full effect of the external fields is included in the time-dependent real control parameter $\lambda(t)$. The effective energy level diagram related to the environment-induced transitions for this model system is schematically presented in Fig.~\ref{fig:model}.

The slow driving is accounted for by the time-dependent basis transformations which shift the transition frequencies and alter the corresponding dissipative rates. Note that the transformation scheme is only limited by the assumptions on (super)adiabaticity ($\alpha_n \ll 1 $), the system--environment coupling strength, and the environment autocorrelation time. Thus, the transformation scheme allows for an arbitrary drive strength as long as the time derivatives of the drive are small enough to guarantee $\alpha_n \ll 1 $. The model system describes, for instance, a spin-1/2 particle in a time-dependent magnetic field $\vec{B}(t)=-[\hbar \omega_0 \vec{z} + 2 \lambda(t)\vec{x}]/(\hbar \gamma)$ where $\gamma$ is the gyromagnetic ratio and $(\vec{x},\vec{y},\vec{z})$ are the Cartesian unit vectors chosen such that the time-independent component of the magnetic field is along $z$. The model is applicable beyond spin systems, and the theory derived retains its validity provided that the two-level approximation is satisfied under the action of the external drive, i.e., the two-level approximation does not essentially distort the instantaneous energy spacing of the system. A notable example of non-applicability is the harmonic oscillator driven with one of its canonical variables, for which the above two-level approximation produces incorrect eigenenergies.

We assume the following form of the coupling operator:
\begin{equation}
\hat{Y} = g^* \hat{\sigma}_++ g \hat{\sigma}_-,
\end{equation}
where $g$ denotes the complex coupling strength related to the two-level operations. In this paper, we adopt the convention of $g$ carrying the units of energy implying that $\hat{X}$ is dimensionless. We choose our two-state model system to mathematically facilitate a comparison~\footnote{Typically the interaction of a photonic bath with a quantum system is described by $\hat{X} = \sum_{\mu} (d_{\mu}^{*} \hat{b}_{\mu}^{\dagger} + d_{\mu} \hat{b}_{\mu})$, where $\hat{b}_{\mu}^{\dagger}$ ($\hat{b}_{\mu}$) creates (annihilates) a photon at frequency $\omega_{\mu}$ and $d_{\mu}$ is a dimensionless scalar describing the interaction related to the $\mu$th mode. An application of the rotating wave approximation results in $\hat{V} = \sum_{\mu} (c_{\mu}\hat{\sigma}_+\hat{b}_{\mu} + c_{\mu}^*\hat{\sigma}_-\hat{b}_{\mu}^{\dagger})$, where $c_{\mu} = g^*d_{\mu}$, corresponding to the system--bath coupling operator used in Ref.~\cite{prl111/093602}.} with Ref.~\cite{prl111/093602}, which proposes a physical realization of the system as a superconducting device~\cite{prl89/117901, prl93/187003} monitored by calorimetry of the coupled environment~\cite{njp15/115006}.

The instantaneous eigenenergies of $\hat{H}_S(t)$ take the form
\begin{equation}
E_{e/g}^{(1)} = \pm\frac{\sqrt{(\hbar\omega_0)^2+4\lambda^2}}{2}.
\label{eq:E1}
\end{equation}
The corresponding adiabatic states are $\ket{e^{(1)}} = C_{ee}^{(1)}\ket{e} + C_{eg}^{(1)}\ket{g}$ and $\ket{g^{(1)}} = C_{ge}^{(1)}\ket{e} + C_{gg}^{(1)}\ket{g}$, where
\begin{equation}
\begin{split}
C_{eg}^{(1)}& =\left[ 1+\frac{1}{4\lambda^2}\left(  \hbar\omega_0+ \sqrt{(\hbar\omega_0)^2+4\lambda^2}   \right)^2 \right]^{-1/2}, \\
C_{ee}^{(1)}& =C_{eg}^{(1)}\frac{1}{2\lambda}\left(  \hbar\omega_0+ \sqrt{(\hbar\omega_0)^2+4\lambda^2}   \right), \\
C_{gg}^{(1)}& =\left[ 1+\frac{1}{4\lambda^2}\left(  \hbar\omega_0- \sqrt{(\hbar\omega_0)^2+4\lambda^2}   \right)^2 \right]^{-1/2}, \\
C_{ge}^{(1)}& =C_{gg}^{(1)}\frac{1}{2\lambda}\left(  \hbar\omega_0-\sqrt{(\hbar\omega_0)^2+4\lambda^2}   \right).
\end{split}
\label{eq:Cadi}
\end{equation}
Note that these states explicitly define the first coordinate transformation as $\hat{D}_1 = \ket{e^{(1)}}\bra{e} + \ket{g^{(1)}}\bra{g}$. In Eqs.~(\ref{eq:E1})--(\ref{eq:Cadi}) and in the following, we omit explicitly marking the time-dependence for each variable for notational convenience.

As an example of the renormalization procedure, we present the first-order superadiabatic states by defining $w_{mn}^{(1)} = \braket{m|\hat{w}_1|n} = -i\braket{m|\hat{D}_1^{\dagger}\dot{\hat{D}}_1|n}$, whose two-level components can be solved directly from Eq.~(\ref{eq:Cadi}). In fact, utilizing the instantaneous eigenproblem allows us to write
\begin{equation}
w_{mn}^{(1)} = -i\frac{\langle m^{(1)}|\frac{\partial \hat{H}_S}{\partial \lambda}\dot{\lambda}|n^{(1)}\rangle}{E_n^{(1)}-E_m^{(1)}},
\end{equation}
for $m\neq n$. In our case, this component takes the form
\begin{equation}
w_{ge}^{(1)}  = -i\dot{\lambda}\frac{[C_{ee}^{(1)}(C_{gg}^{(1)})^*+C_{eg}^{(1)}(C_{ge}^{(1)})^*]}{E_e^{(1)}-E_g^{(1)}},
\label{eq:wge1}
\end{equation}
and it can be shown using Eq.~(\ref{eq:Cadi}) that $w_{eg}^{(1)} = (w_{ge}^{(1)})^* = - w_{ge}^{(1)}$ and $w_{ee}^{(1)} = w_{gg}^{(1)} = 0$. These identities are due to our convenient phase selection for the adiabatic states, and correspond to the optimal phase selection method introduced in Ref.~\cite{pra82/062112}, based on tracking the Berry connection~\cite{prsla392/45} in order to minimize the local adiabatic parameter at all times. However, we expect that the effect of the method on the reduced dynamics is weak compared to that of the coordinate transformations in the dissipative approach that we adopt in the following.

We can solve the eigenproblem $[\hat{\tilde{H}}^{(1)}_S(t) + \hbar \hat{w}_1(t)]\ket{\tilde{m}^{(2)}(t)} = E_m^{(2)}(t)\ket{\tilde{m}^{(2)}(t)}$ to obtain
\begin{equation}
E^{(2)}_{e/g} = \pm \frac{1}{2}\sqrt{\left(E_e^{(1)}-E_g^{(1)}\right)^2 + 4\hbar^2 |w_{ge}^{(1)}|^2},
\label{eq:E2}
\end{equation}
where the quantities defined in Eqs.~(\ref{eq:E1}) and (\ref{eq:wge1}) are used. The corresponding eigenstates are $\ket{\tilde{e}^{(2)}} = C_{ee}^{(2)}\ket{e} + C_{eg}^{(2)}\ket{g}$ and $\ket{\tilde{g}^{(2)}} = C_{ge}^{(2)}\ket{e} + C_{gg}^{(2)}\ket{g}$, where
\begin{equation}
\begin{split}
C_{eg}^{(2)}& =\left[ 1+\frac{1}{|\hbar w_{ge}^{(1)}|^2}\left(  E^{(2)}_e-E^{(1)}_g \right)^2 \right]^{-1/2}, \\
C_{ee}^{(2)}& =C_{eg}^{(2)}\frac{1}{\hbar w_{ge}^{(1)}} \left(  E^{(2)}_e-E^{(1)}_g \right), \\
C_{gg}^{(2)}& =\left[ 1+\frac{1}{|\hbar w_{ge}^{(1)}|^2}\left(  E^{(2)}_g-E^{(1)}_g \right)^2 \right]^{-1/2}, \\
C_{ge}^{(2)}& =C_{gg}^{(2)}\frac{1}{\hbar w_{ge}^{(1)}}\left(  E^{(2)}_g-E^{(1)}_g \right).
\end{split}
\label{eq:C1sup}
\end{equation}
Note that the diagonal parts of $\hat{w}_1$ do not appear in Eqs.~(\ref{eq:E2}) and (\ref{eq:C1sup}) due to the optimal phase selection method described above. Hence, the second unitary transformation is defined as $\hat{D}_2 = \ket{\tilde{e}^{(2)}}\bra{e} + \ket{\tilde{g}^{(2)}}\bra{g}$, and the first-order superadiabatic states are $\ket{e^{(2)}} = \hat{D}_1\hat{D}_2 \ket{e} = [C_{ee}^{(2)}C_{ee}^{(1)} + C_{eg}^{(2)}C_{ge}^{(1)}]\ket{e} + [C_{ee}^{(2)}C_{eg}^{(1)} + C_{eg}^{(2)}C_{gg}^{(1)}]\ket{g}$ and $\ket{g^{(2)}} = \hat{D}_1\hat{D}_2 \ket{g} = [C_{ge}^{(2)}C_{ee}^{(1)} + C_{gg}^{(2)}C_{ge}^{(1)}]\ket{e} + [C_{ge}^{(2)}C_{eg}^{(1)} + C_{gg}^{(2)}C_{gg}^{(1)}]\ket{g}$. By using the eigenproblem defined above, we have for $w_{mn}^{(2)} = -i\braket{m|\hat{D}_2^{\dagger}\dot{\hat{D}}_2|n}$ the identity
\begin{equation}
w_{mn}^{(2)} = -i \frac{\braket{\tilde{m}^{(2)}|\partial_t (\hat{\tilde{H}}_S^{(1)}+\hbar \hat{w}_1)|\tilde{n}^{(2)}}}{E_n^{(2)}-E_m^{(2)}},
\label{eq:wge2}
\end{equation}
for $m\neq n$. For our two-level system, this results in
\begin{equation}
\begin{split}
w_{ge}^{(2)} = -i \{ &(C_{ge}^{(2)})^* [C_{ee}^{(2)}\dot{E}_e^{(1)}-\hbar C_{eg}^{(2)} \dot{w}_{ge}^{(1)}] \\ + &(C_{gg}^{(2)})^* [C_{eg}^{(2)} \dot{E}_g^{(1)} + \hbar C_{ee}^{(2)} \dot{w}_{ge}^{(1)} ] \} \\ \times & \frac{1}{E_e^{(2)}-E_g^{(2)}},
\end{split}
\label{eq:wge2s}
\end{equation}
where we used $(w_{ge}^{(1)})^* = - w_{ge}^{(1)}$. Using Eqs.~(\ref{eq:E1}), (\ref{eq:Cadi}), and (\ref{eq:wge1}), the time derivatives can be expressed in terms of the control parameter as
\begin{equation}
\begin{split}
\dot{E}_{e/g}^{(1)} = \pm \dot{\lambda} \frac{2\lambda}{\sqrt{4\lambda^2+\hbar^2\omega_0^2}},
\end{split}
\label{eq:derE}
\end{equation}
and
\begin{equation}
\begin{split}
\dot{w}_{ge}^{(1)} = -i\ddot{\lambda} \frac{\hbar \omega_0 \lambda}{4\lambda^3+\hbar^2\omega_0^2\lambda} + i(\dot{\lambda})^2 \frac{8\hbar \omega_0 \lambda}{(4\lambda^2+\hbar^2\omega_0^2)^2}.
\end{split}
\label{eq:derw}
\end{equation}
Equation (\ref{eq:C1sup}) implies that $C_{eg}^{(2)}, C_{gg}^{(2)} \in \mathbb{R}$ and $iC_{ee}^{(2)}, iC_{ge}^{(2)} \in \mathbb{R}$ which yield $w_{eg}^{(2)} = (w_{ge}^{(2)})^* = - w_{ge}^{(2)}$. We can additionally show using Eq.~(\ref{eq:C1sup}) that $w_{ee}^{(2)} = w_{gg}^{(2)} = 0$ corresponding to the optimal phase selection for the first-order superadiabatic states~\cite{pra82/062112}. The higher-order superadiabatic bases are accessed by continuing the procedure. To allow for an analytical treatment up to the first superadiabatic basis, we additionally write $m_1^{(1)} = 2 \text{Re}\lbrace g(C_{gg}^{(1)})^*C_{ge}^{(1)}\rbrace$, $m_2^{(1)} = g(C_{gg}^{(1)})^*C_{ee}^{(1)}+g^* (C_{ge}^{(1)})^*C_{eg}^{(1)}$, $m_1^{(2)} = (|C_{gg}^{(2)}|^2-|C_{ge}^{(2)}|^2) m_1^{(1)} + 2\mathrm{Re} \{C_{ge}^{(2)}(C_{gg}^{(2)})^* m_2^{(1)} \}$ and $m_2^{(2)} = [(C_{gg}^{(2)})^* C_{eg}^{(2)} - (C_{ge}^{(2)})^* C_{ee}^{(2)}] m_1^{(1)}  + (C_{gg}^{(2)})^* C_{ee}^{(2)} m_2^{(1)} + (C_{ge}^{(2)})^* C_{eg}^{(2)} (m_2^{(1)})^*$, where the state amplitudes are given in Eqs.~(\ref{eq:Cadi}) and (\ref{eq:C1sup}).

\begin{figure}
\centering
\includegraphics[width=0.45\textwidth]{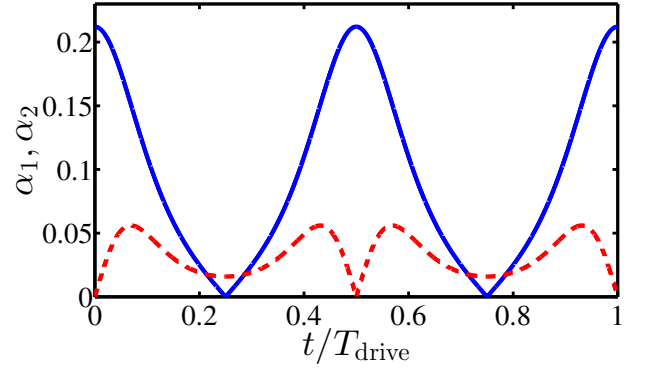}
\caption{(Color online) Adiabatic (solid line) and first superadiabatic (dashed line) parameters during the driving cycle. We assume that $\lambda_0=\hbar\omega_0/2$ and $\omega_d=3\omega_0/10$. The length of the sinusoidal driving cycle is defined by $T_{\mathrm{drive}}=2\pi/\omega_d$.}
\label{fig:par1}
\end{figure}

Using Eqs.~(\ref{eq:wge1}) and (\ref{eq:wge2s}), the (super)adiabatic parameter $\alpha_n = ||\hat{w}_n||/\omega_{01}^{(n)}$  can be evaluated for $n=1,2$ to analyze the adiabaticity of the driving cycle and the accuracy of Eq.~(\ref{eq:ME}) with regards to the selected driving protocol. We assume that the system is under strong sinusoidal driving such that
\begin{equation}
\lambda(t) = \lambda_0 \sin (\omega_d t),
\end{equation}
where $\lambda_0$ and $\omega_d$ denote the amplitude and angular frequency corresponding to the prescribed protocol, respectively. The adiabatic and first superadiabatic parameters for the selected driving cycle are shown in Fig.~\ref{fig:par1}.
They indicate that the error in the master equation (Eq. \eqref{eq:ME}) stemming from the remaining inaccuracy in the selection of the dynamical basis is mitigated by the renormalization procedure up to the first superadiabatic basis. Such mitigation is generally true for further basis rotations. The closer the system is to the adiabatic limit~\cite{prsla392/45}, the more rotations can be carried out to improve accuracy.

\section{Quantum trajectories and work} \label{sec:Qtraj}

To access the distribution of work injected into the open quantum system during the driving protocol, we utilize the so-called {\it quantum jump method}, also referred to as the quantum stochastic wave function method or the Monte Carlo wave function technique~\cite{prl68/580, pra45/4879, pra46/4363, OSAQO, rmp70/101}. The method is based on unraveling the quantum evolution given by a master equation in the Lindblad form into stochastic trajectories describing single realizations of the dissipative dynamics~\cite{ap248/95, EQ}. Each trajectory represents a pure state evolution of the open quantum system under continuous measurements of its environment, corresponding to either nonunitary time evolution of the state vector or its instantaneous transitions referred to as quantum jumps. Consequently, the continuous measurements allow for in situ monitoring of the system state~\cite{pra47/1652}. For each trajectory, work is obtained as a combination of applying the two-measurement approach (TMA)~\cite{arxiv/0007360, rmp81/1665, rmp83/771} to extract the internal energy of the system $E_S(t)$, and allocating to each quantum jump event $j$ the corresponding heat transferred to the environment $Q(t_j)$. In accordance with the first law of thermodynamics, the trajectory-dependent work performed on the system during a driving protocol $t_{\mathrm{init}} \rightarrow t_{\mathrm{final}}$ is $W = \Delta E_S + Q_{\mathrm{tot}}$, where $\Delta E_S = E_S(t_{\mathrm{final}})-E_S(t_{\mathrm{init}})$, and the total heat transferred to the environment is $Q_{\mathrm{tot}} = \sum_j Q(t_j)$, where the summation is over all jump events taking place along the single trajectory.

Let us denote the state vector of the open system by $\vert \phi(t) \rangle$. The non-Hermitian effective Hamiltonian yielding the time-evolution during a jump-free interval~\cite{prl68/580, pra45/4879, pra46/4363, OSAQO, rmp70/101} is defined by
\begin{equation}
\hat{H}_{\mathrm{eff}}^{(n)}(t) = \hat{H}_S(t)-\frac{i \hbar}{2} \sum_{i=0}^2 \hat{L}_{(n,i)}^\dagger(t) \hat{L}_{(n,i)}(t),
\label{eq:HeffQJ}
\end{equation}
where the Lindblad operators are given in Eq.~(\ref{eq:L}). The corresponding nonunitary effective time evolution operator is
\begin{equation}
\hat{U}_{\mathrm{eff}}^{(n)}(t_2,t_1) = \mathcal{T} \exp \left\{-\frac{i}{\hbar} \int_{t_1}^{t_2} \hat{H}_{\mathrm{eff}}^{(n)}(\tau)d\tau\right\},
\end{equation}
where $\mathcal{T}$ is the time ordering. We define a small time step $\Delta t$ during which each event takes place and, correspondingly, the evolved state can be accurately approximated by
\begin{equation}
\ket{\phi (t+\Delta t)} = \frac{\hbar\hat{I}_S - i\Delta t \hat{H}_{\mathrm{eff}}^{(n)}(t)}{\hbar N_{\phi}(t+\Delta t)}\ket{\phi(t)},
\end{equation}
where $N_{\phi}(t+\Delta t) = ||[1-i\Delta t \hat{H}_{\mathrm{eff}}^{(n)}(t)/\hbar]\ket{\phi(t)}||$ accounts for the normalization of the state, and $\hat{I}_S$ is the identity operator acting in the Hilbert space of the two-level system. The probability for a jump event to occur during the same time interval is given by $p_n(t) = \sum_{i=0}^2 p_{(n,i)}(t)$, where $p_{(n,i)}(t)  = \Delta t \braket{ \phi(t) | \hat{L}_{(n,i)}^\dagger(t) \hat{L}_{(n,i)} (t) | \phi(t)}$ accounts for the contribution of the $i$th Lindblad operator~\cite{prl68/580, pra45/4879, pra46/4363, OSAQO, rmp70/101}. If a jump event occurs, it takes place along the $i$th channel with probability $p_{(n,i)}(t)/p_n(t)$ and the system collapses after the jump to the normalized state
\begin{equation}
\ket{\phi(t+\Delta t)} = \frac{\hat{L}_{(n,i)}}{\sqrt{p_{(n,i)}/ \Delta t}}  \ket{\phi(t)}. \label{eq:col}
\end{equation}

The number of basis rotations directly affects both the jump probabilities as well as the states after the jump and no-jump events. The physical framework related to each realization is best understood using the $n$th (super)adiabatic basis. The second term in Eq.~(\ref{eq:HeffQJ}) for $\hat{H}_{\mathrm{eff}}^{(n)}(t)$ is anti-Hermitian and diagonal in this basis, causing environment-induced population transfer between the basis states, whereas $\hat{H}_S(t)$ accounts for all direct driving-induced transitions between the states. Each jump event corresponds to either an instantaneous collapse [see Eqs. \eqref{eq:L} and \eqref{eq:col}] to the $n$th ground state [$\hat{L}_{(n,0)}$], an instantaneous collapse to the $n$th excited state [$\hat{L}_{(n,1)}$], or a phase flip [$\hat{L}_{(n,2)}$] within this basis. This framework extends the approach of Ref.~\cite{prl111/093602}, where the effect of driving is omitted in the formulation of the dissipator. Such an approach can be considered to be formally equal to the level of recursion $n=0$ such that $\hat{\tilde{H}}_S^{(0)}(t) = \hat{H}_S(t')$ and $\hat{\tilde{H}}_S^{(0)} \ket{k^{(0)}} = E_k^{(0)} \ket{k^{(0)}}$, where $t'$ is an arbitrary fixed time instance. Selection of $t'$ such that $\lambda(t')=0$ corresponds to the treatment of Ref.~\cite{prl111/093602} for the two-level system. Here, the assumption is that the driving is weak enough for its effect on the environmental transitions to be negligible. This implies that the transition rates are determined by a constant energy gap $\hbar \omega_0$ with each jump projecting the state to an eigenstate of $\hat{\sigma}_z$. In addition, the dephasing channel in the quantum jump method was not considered in Ref.~\cite{prl111/093602}.

The change in the trajectory-dependent internal energy is defined using the TMA through projective measurements in the eigenbasis of $\hat{H}_S(t)$ such that $E_S(t_{\mathrm{init}}) = E_k^{(1)}(t_{\mathrm{init}})$ and $E_S(t_{\mathrm{final}}) = E_l^{(1)}(t_{\mathrm{final}})$, where $k$ and $l$ denote which eigenenergy was obtained at the initial and final time instances, respectively. In addition, we denote the probability distribution of the initial measurement results as $P_{\mathrm{init}}[E_S(t_{\mathrm{init}})] = \sum_m P[E_m^{(1)}(t_{\mathrm{init}})] \delta [E_S(t_{\mathrm{init}}) - E_m^{(1)}(t_{\mathrm{init}})]$, where $P[E_m^{(1)}(t_{\mathrm{init}})]$ denotes the probability of measuring the $m$th eigenenergy. Within the present quantum jump method, the heat transfer associated to a jump at $t=t_j$ is determined by the Lindblad operators in Eq.~(\ref{eq:L}), and given by $Q^{(n)}(t_j) = \hbar \omega_{01}^{(n)}(t_j)$ if the jump occurs along the $i=0$ channel, by $Q^{(n)}(t_j) = - \hbar \omega_{01}^{(n)}(t_j)$ if it occurs along the $i=1$ channel, and by $Q^{(n)}(t_j) = 0$ if it occurs along the $i=2$ channel. Note that even though no heat transfer occurs along the dephasing channel, it still contributes to the stochastic dynamics and should be included. The trajectory-dependent work corresponding to the $n$th-order dynamics is given by
\begin{equation}
W^{(n)} = \Delta E_S + Q^{(n)}_{\mathrm{tot}},
\label{eq:Wn}
\end{equation}
where the total heat is defined by $Q^{(n)}_{\mathrm{tot}} = \sum_j Q^{(n)}(t_j)$. The $n$-dependence of the stochastic dynamics is naturally inherited by the trajectory-dependent work through the heat exchange term. As the description of the dissipative dynamics becomes more accurate with increasing $n$, so does the heat allocated to each jump event. Consequently, the total heat $Q^{(n)}_\textrm{tot}$ approaches the physically observable $n$-independent heat.

Equation~(\ref{eq:ME}) is in the Lindblad form~\cite{cmp48/119} for each $n$. Additionally, the corresponding Lindblad operators defined in Eq.~(\ref{eq:L}) satisfy a local detailed balance condition~\cite{PhysRevE.85.031110,NJP15/085028} as $\hat{L}_{(n,0)}\propto \hat{L}_{(n,1)}^\dagger$ and $\hat{L}_{(n,2)}=\hat{L}_{(n,2)}^\dagger$. Thus, the stochastic dynamics defined above can be shown to fulfill the integral fluctuation theorem such that~\footnote{See Appendix~\ref{app:IFT} for a full derivation. Similar results have previously been obtained in Refs. \cite{PhysRevE.85.031110,NJP15/085028}.} 
\begin{equation}
\Braket{  \frac{\bar{P}[\bar{E}_l^{(1)}(t_{\mathrm{init}})]}{P[E_k^{(1)}(t_{\mathrm{init}})]} \prod_{j=1}^N \frac{\bar{\Gamma}_{(n,m_j)}(\bar{t}_j)}{\Gamma_{(n,m_j)}(t_j)} }_{(n)} = 1,
\label{eq:RI}
\end{equation}
where the multiplication is over all jump events occurring along a single $N$-jump trajectory, $m_j$ denotes the channel along which the $j$th jump event takes place, the notation $\bar{a}$ refers to the value taken by arbitrary $a$ during the traversal of a time-reversed trajectory~\cite{rmp81/1665}, and $\braket{\cdots}_{(n)}$ denotes an ensemble average over all the possible trajectories generated by the $n$th-order stochastic dynamics. In particular, $\bar{P}[\bar{E}_l^{(1)}(t_{\mathrm{init}})]$ is the probability to obtain $\bar{E}_l^{(1)}(\bar{t}) = E_l^{(1)}(t)$ from the projective measurement taking place at the initial time $\bar{t} = t_{\mathrm{init}}$ of the time-reversed evolution. The transition rates corresponding to the time-reversed dynamics are given by $\bar{\Gamma}_{(n,0)}(\bar{t}_j)=\Gamma_{(n,1)}(t_{j})$, $\bar{\Gamma}_{(n,1)}(\bar{t}_j)=\Gamma_{(n,0)}(t_{j})$, and $\bar{\Gamma}_{(n,2)}(\bar{t}_j)=\Gamma_{(n,2)}(t_{j})$.
Note that the temporal variables are simply connected by $\bar{t} = (t_{\mathrm{init}} + t_{\mathrm{final}}) - t$, e.g, $\bar{t}_j = (t_{\mathrm{init}} + t_{\mathrm{final}}) - t_j$.
Because $\bar{\Gamma}_{(n,2)}(\bar{t}_j)/\Gamma_{(n,2)}(t_{j}) = 1$, the dephasing channel does not directly contribute to the trajectory-dependent product in Eq.~(\ref{eq:RI}) and, hence, only affects the left-hand side through the dynamics of the state, that is, within the confines of the ensemble averaging.

Let us assume that the system is initially in thermal equilibrium at temperature $T$ such that $P[E_m^{(1)}(t_{\mathrm{init}})] = e^{-\beta E_m^{(1)}(t_{\mathrm{init}})}/Z(t_{\mathrm{init}})$, where $\beta = 1/(k_B T)$, $k_B$ denotes the Boltzmann constant, and $Z(t) = \mathrm{Tr}_S \{ e^{-\beta \hat{H}_S(t)} \}$ is the system partition function. If we select the initial state of the time-reversed evolution such that $\bar{P}[\bar{E}_l^{(1)}(t_{\mathrm{init}})] = e^{-\beta E_l^{(1)}(t_{\mathrm{final}})}/Z(t_{\mathrm{final}})$, it corresponds to the same equilibration by the heat bath as that prior to the original evolution, and Eq.~(\ref{eq:RI}) takes the form
\begin{equation}
\Braket{ e^{-\beta[E_l^{(1)}(t_{\mathrm{final}}) - E_k^{(1)}(t_{\mathrm{init}})]} \prod_{j=1}^N \frac{\bar{\Gamma}_{(n,m_{j})}(\bar{t}_j)}{\Gamma_{(n,m_{j})}(t_j)} }_{(n)} =e^{-\beta\Delta F},
\label{eq:RI2}
\end{equation}
where $\Delta F=F(t_{\mathrm{final}}) - F(t_{\mathrm{init}}) = -\beta^{-1} \ln [Z(t_{\mathrm{final}})/Z(t_{\mathrm{init}})]$ is the free energy difference between a reference equilibrium state described by the Hamiltonian $\hat{H}_S(t_\mathrm{final})$ and the initial equilibrium state. The above-described properties of the time-reversed evolution yield $\bar{\Gamma}_{(n,m_{j})}(\bar{t}_j)/\Gamma_{(n,m_{j})}(t_{j}) = e^{-\beta Q^{(n)}(t_{j})}$ for each dissipative channel, assuming that the noise source follows detailed balance $S(\omega_{01}^{(n)}) = e^{\beta \hbar \omega_{01}^{(n)}} S(-\omega_{01}^{(n)})$ at all times. Using this definition, Eq.~(\ref{eq:RI2}) yields the Jarzynski equality~\cite{prl78/2690} for nearly adiabatic driving
\begin{equation}
\braket{e^{-\beta W^{(n)}}}_{(n)} =e^{-\beta\Delta F}.
\label{eq:JE}
\end{equation}
Notably, this identity always holds when the $n$th-order definition for the trajectory-dependent work is consistently used in association with the $n$th-order stochastic dynamics. However, a typical experimental probe of the dynamics is only able to determine the jump time distribution~\cite{prl109/180601, np9/644, prl113/030601, pnas111/13786} and not directly the heat transfer associated to each jump. In this case, the evaluation of the work distribution must externally impose to each detected jump event the corresponding heat transfer and, accordingly, adopt a definition for the trajectory-dependent work. Depending on the adopted definition, the Jarzynski equality may or may not be accurately retrieved using the experimental jump time distribution. For nearly adiabatic driving, this means that using a low-order definition for the work, can potentially result in significant error in determining the free energy difference even if the trajectories are accurately traced by the experiment.

\section{Work statistics for nearly adiabatic driving} \label{sec:num}

In the following numerical calculations, we assume that the sinusoidally driven system is initially in thermal equilibrium, and that the noise spectrum is ohmic such that $S(\omega)=2 \mu \omega /[\hbar(1-e^{- \beta \hbar \omega})]$~\cite{rmp82/1155}, where $\mu$ is the damping constant related to the noise source. This spectral density fulfills the detailed balance condition and is also applicable to a variety of physical cases. Furthermore, we introduce dephasing by assuming that $S(0) = 2\mu k_BT_0/\hbar^2$, where $T_0$ is the effective dephasing temperature.

\begin{figure}[b] \centering
\includegraphics[width=0.48\textwidth]{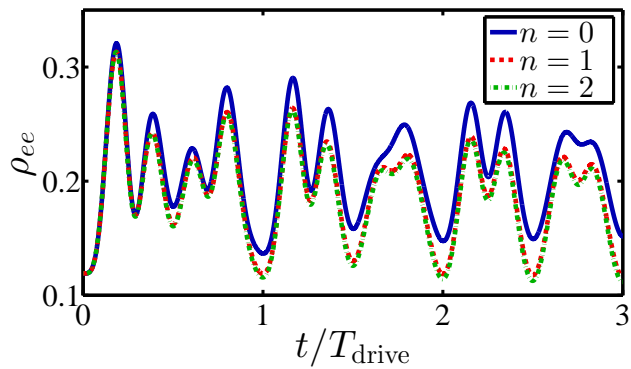}
\caption{(Color online) Population of the excited diabatic state, $\rho_{ee}$, using dynamics calculated with $n$ basis rotations in the adiabatic renormalization scheme. The length of a single sinusoidal cycle is $T_{\mathrm{drive}} = 2\pi/\omega_d$. The parameters are selected such that $\omega_d = 3\omega_0/10$, $\lambda_0 = \hbar \omega_0/2$, $g=\hbar \omega_0/(5 \sqrt{2})$, $T=\hbar \omega_0/(2k_B)$, $T_0= 2T$, and $\mu = 1/(\hbar \omega_0^2)$. The system is initialized to the Gibbs state corresponding to the above bath temperature. The driving protocol is discretized using $10^7$ equidistant time steps and the number of trajectories is $10^7$.}
\label{fig:rho}
\end{figure}
The numerical calculations are carried out for the first three orders of basis transformations in the adiabatic renormalization scheme, i.e., $n=0,1,2$. The calculations are performed using the master equation of Eq. \eqref{eq:ME} in the corresponding $n$th-order superadiabatic basis.  The parameters are selected such that the angular frequency and amplitude of the sinusoidal driving protocol are $\omega_d = 3\omega_0/10$ and $\lambda_0 = \hbar \omega_0/2$, respectively, the system--environment coupling strength is $g=\hbar \omega_0/(5 \sqrt{2})$, the effective bath and dephasing temperatures are $T=\hbar \omega_0/(2k_B)$ and $T_0= 2 T = \hbar \omega_0/k_B$, and the damping constant is $\mu = 1/(\hbar \omega_0^2)$ given in the units of inverse power corresponding to the unit convention adopted above. The dynamics are recorded over three consecutive drive cycles such that the initial state is the thermal equilibrium state corresponding to the above bath temperature. With these system parameters, the local superadiabatic parameter $\alpha_2$ is relatively small, as shown in Fig.~\ref{fig:par1}, such that the superadiabatic results provide a faithful approximation of the real dynamics.
\begin{figure}[b]
\centering
\includegraphics[width=0.48\textwidth]{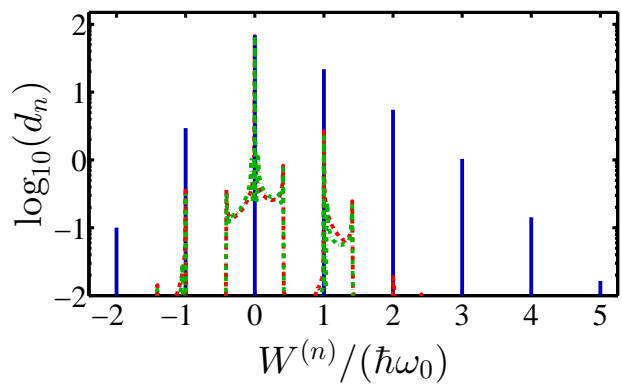}
\caption{(Color online) Probability density of work $d_n(W^{(n)})$ using different numbers of basis rotations, $n$, in the adiabatic renormalization scheme. The solid line (blue) corresponds to $n=0$, the dashed line (red) to $n=1$, and the dash-dotted line (green) to $n=2$. The density function is calculated with a bin size $10^{-2}/(\hbar \omega_0)$ and the parameters are selected to be the same as in Fig.~\ref{fig:rho}.}
\label{fig:W}
\end{figure}
The effect of the basis rotations to the dynamics is illustrated in Fig.~\ref{fig:rho}, where the population of the excited diabatic state $\rho_{ee}=\braket{e |\hat{\rho}_S |e}$ is calculated using different numbers of basis rotations ($n=0,1,2$) in the renormalization scheme.

Although all approximative orders of dynamics closely follow each other at the beginning of the drive, both the accumulation of the nonadiabatic corrections and the discrepancy between the bases in which dissipation acts, result in a clear difference in the populations during the three drive cycles. This difference is especially pronounced between the diabatic basis and the higher-order bases due to relatively slow driving, implying that $n=1$ provides a rather accurate approximation of the dynamics. Based on Fig.~\ref{fig:rho}, $n=3$ would not provide a significant correction to the population dynamics.

In Fig.~\ref{fig:W}, we present the probability distributions of work calculated numerically using a finite number of trajectories [see Appendix~\ref{app:pnw} for the exact definition of the probability densities of work, $d_n(W^{(n)})$]. The distributions are calculated utilizing different numbers of basis rotations to describe the interplay between driving and dissipation.
\begin{figure}[t]
\centering
\includegraphics[width=0.48\textwidth]{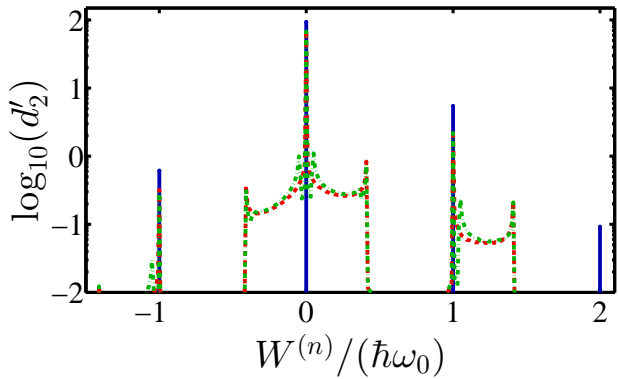}
\caption{(Color online) Probability density of work $d'_2(W^{(n)})$ using an $n$th-order work variable while fixing the dynamics to $n=2$. The solid line (blue) corresponds to $n=0$, the dashed line (red) to $n=1$, and the dash-dotted line (green) to $n=2$. The density function is calculated with a bin size $10^{-2}/(\hbar \omega_0)$ and the parameters are selected to be the same as in Fig.~\ref{fig:rho}.}
\label{fig:W_sa_dynamics}
\end{figure}
Note that both the dynamics and the trajectory-dependent work $W^{(n)}$ depend on the number of transformations carried out. For $n=0$, the trajectory-dependent work can only obtain discrete value as multiples of $\hbar \omega_0$ indicated by the relevent distribution in Fig.~\ref{fig:W} consisting of $\delta$ peaks. Using higher-order dynamics enables the time-dependent driving to be accounted for in the environment--induced transitions and, consequently, the probability distributions become continuous in work as evident from Fig.~\ref{fig:W}. This continuity is a fundamental consequence of driving and not caused by nonadiabatic transitions due to the selection of a nonvanishing driving frequency. In the adiabatic limit, the direct driving--induced transitions vanish from the jump-free evolution but the time-dependence of the exchanged heat for each transition event still remains.
\begin{table*}
\begin{tabular}{| c | c | c | c |} \hline
& $n=2$  & $n=1$  & $n=0$  \\ \hline
$ \bigg. \left| 1-\langle e^{-\beta W^{(n)}} \rangle_{(n)} \right|$ & $(1\pm 3)\times 10^{-4}$ & $(0\pm 4)\times 10^{-4}$ & $(5\pm 7)\times 10^{-4}$ \\ \hline
$\bigg. \langle  W^{(n)} \rangle_{(n)}/(\hbar \omega_0)$  & $(693\pm 1)\times 10^{-4}$ & $(969\pm 2)\times 10^{-4}$ & $(3335 \pm 1)\times 10^{-4}$ \\ \hline
$\bigg. \langle (W^{(n)})^2 \rangle_{(n)}/(\hbar \omega_0)^2$  & $(952\pm 2)\times 10^{-4}$ & $(1373\pm 2)\times 10^{-4}$ & $(5915\pm 3)\times 10^{-4} $\\ \hline \hline
$ \bigg. \left| 1-\langle e^{-\beta W^{(n)}} \rangle_{(2)} \right|$  & $(1 \pm 9) \times 10^{-5}$ & $(11 \pm 9) \times 10^{-5}$ & $(867 \pm 6) \times 10^{-5}$ \\ \hline
$\bigg. \langle  W^{(n)} \rangle_{(2)}/(\hbar \omega_0)$  & $(6923 \pm 3) \times 10^{-5}$   & $(6901 \pm 3) \times 10^{-5}$  & $(5055 \pm 3) \times 10^{-5}$ \\ \hline
$\bigg. \langle (W^{(n)})^2 \rangle_{(2)}/(\hbar \omega_0)^2$  &$(9507 \pm 4) \times 10^{-5}$ &$(9450 \pm 4) \times 10^{-5}$ & $(6473 \pm 3) \times 10^{-5}$ \\
\hline
\end{tabular}
\caption{The exponentiated work and the first two moments of work using its $n$th-order definition in conjunction with the dynamics of equal order (top rows) as well as that of $n=2$ (bottom rows). The top rows are calculated with $10^7$ timesteps, $2 \times 10^7$ trajectories for $n=1,2$ and $10^8$ trajectories for $n=0$. The bottom rows are calculated with $10^6$ timesteps and $3 \times 10^8$ trajectories. The statistical uncertainty arising from averaging over the trajectories is estimated using the standard error of the mean, $\sigma$. In the table, the error estimates allow a maximum of $1.96\sigma$ deviation corresponding to a $95\%$ confidence interval of the mean. The system parameters are selected to be the same as in Fig.~\ref{fig:rho}.}
\label{tab:averages}
\end{table*}

The asymmetry of the distributions in Fig.~\ref{fig:W} with respect to $W^{(n)}=0$ is caused by two factors: the selection of the initial state as the Gibbs distribution and the assertion of a noise source fulfilling detailed balance. This is embodied by the Jarzynski equality in Eq.~(\ref{eq:JE}) setting a definite condition for the resulting distribution. For a closed loop in the control parameter space, $\Delta F = 0$ implying that $e^{-\beta W^{(n)}}$ weighted by the probability density must integrate to unity and, hence, more density must always be found on the positive work values. To emphasize the necessity of defining the work variable in an accurate manner, we present probability densities of work $d'_2(W^{(n)})$ such that the dynamics are calculated using $n=2$ but the work variable uses $n=0,1,2$ in Fig.~\ref{fig:W_sa_dynamics}. In other words, $d'_2(W^{(n)})$ uses $P^{(n)}_{\rightarrow} = P^{(2)}_{\rightarrow}$ for all $n$ in Eq.~(\ref{eq:pnW}). For the nearly adiabatic evolution studied, $d'_2(W^{(1)})$ and $d'_2(W^{(2)})$ are nearly equal, but $d'_2(W^{(0)})$ yields a completely different work distribution, as the zeroth-order approximation for the work variable is unable to capture the time-dependent effect of driving on the heat distribution.

In Table~\ref{tab:averages}, we present the ensemble averaged exponentiated and first two moments of work using $n=0,1,2$. Note that the top rows of the table correspond to values obtained using $d_n(W^{(n)})$ whereas the bottom rows employ $d'_2(W^{(n)})$. 
As predicted by Eq.~(\ref{eq:JE}), we do not observe a statistically significant deviation from the Jarzynski equality when using the consistently computed work distribution $d_n(W^{(n)})$ for each order. Here, we choose $1.96\sigma$ deviation (the 95\% confidence interval) to define the threshold for a statistically significant deviation. 
For $d_n(W^{(n)})$, the moments of work are different for each $n$ due to the differences in the work variables and the probability densities in Fig.~\ref{fig:W}. As expected, the difference between $n=0$ and $n=1$ is much greater than that between $n=1$ and $n=2$.

In Table~\ref{tab:averages}, the work distribution  $d'_2(W^{(n)})$ is consistent with the Jarzynski equality within the statistical uncertainty for $n=2$ guaranteed by Eq.~(\ref{eq:JE}). For $d'_2(W^{(0)})$ however, we observe a clear statistically significant discrepancy with the Jarzynski equality as the average exponentiated work deviates from unity by almost $300\sigma$.  For $d'_2(W^{(1)})$, we also  observe a statistically significant discrepancy with the Jarzynski equality although it is much less pronounced compared with  $d_2'(W^{(0)})$. Due to nearly adiabatic driving, the first two moments of work using $d'_2(W^{(n)})$ deviate less than $1\%$ between $n=1$ and $n=2$, whereas $n=0$ shows greatly different behavior.
\section{Conclusions} \label{sec:con}

We have extended the quantum jump method to nearly adiabatically driven open quantum systems using adiabatic renormalization and derived the required formalism to obtain the stochastic dynamics with respect to the number of renormalization steps. Our framework allows for the quantum jumps to account for the external driving such that both the transition rates and the basis in which the jumps occur become time-dependent. We have constructed the corresponding trajectory-dependent work including the time-dependent heat exchange and derived both the integral fluctuation theorem and the Jarzynski equality for nearly adiabatic driving. Interestingly, it turns out that in this case they both hold even for strong driving as long as the stochastic work variable and dynamics are consistently defined. We expect that a similar framework and identities can be obtained for strong driving beyond the adiabatic limit as the quantum jumps carry the requirement for proper basis selection to any driving speed. We leave such considerations for future work, but note that the shortcut Hamiltonians in {\it transitionless quantum driving}~\cite{jpa42/365303} offer a prominent starting point without the requirement of periodicity in the Floquet theory~\cite{pr304/229}.

We have numerically investigated the work statistics for a nearly adiabatically driven two-level quantum system using different orders of adiabatic renormalization. The level of approximation has a significant effect on both the probability density of work and the corresponding ensemble averaged values. In particular, the diabatic approximation for the dissipative transitions was shown to be unable to capture some of the prominent features emerging in the probability density that arise from driving. These include the fundamental continuity property of the probability density which we observed in the higher orders of the theory. Our observations are potentially critical for future experiments probing only the quantum jump time distributions~\cite{prl109/180601, np9/644, prl113/030601} and not the exchanged heat. To obtain the work statistics, such experiments require one to impose an exchanged heat to each jump event. The conventional scheme using the eigenstates of the undriven Hamiltonian leads to significant error that is potentially visible in, for example, the resulting identification of the free-energy difference between the initial and final reference equilibrium states.

\acknowledgments

The authors thank J. P. Pekola, J. Ankerhold, P. Solinas and E. Aurell for useful discussion. We acknowledge financial support from the V\"ais\"al\"a Foundation and the Academy of Finland through its Centres of Excellence Program under Grant No. 251748 (COMP) and Grants No. 138903, No. 135794, and No. 272806. We have received additional funding from the European Research Council under Starting Independent Researcher Grant No. 278117 (SINGLEOUT) and acknowledge support from the CCQED EU project. I.G.S. acknowledges support from the Government of Russian Federation, Grant 074-U01. The calculations presented above were performed using computer resources within the Aalto University School of Science "Science-IT" project.
\appendix

\section{Secular master equation for nearly adiabatically driven quantum systems} \label{app:der}

In Ref.~\cite{prb84/174507}, a two-level master equation for nearly adiabatic driving is derived using an arbitrary number $n$ of basis rotations in the adiabatic renormalization scheme. Based on the analysis in Ref.~\cite{pra82/062112}, it can be shown that such master equation corresponds to the so-called nonsteered evolution after $(n+1)$ rotations when a linear order approximation in $\alpha_n$ is used for the $(n+1)$th basis states. Following this line of thinking, we assume that $\alpha_{n+1} \ll \alpha_{n}$ and $\hat{\tilde{H}}^{(n+1)} \approx \hat{\tilde{H}}_S^{(n+1)} + \hat{\tilde{V}}^{(n+1)} + \hat{H}_E$  which yields the Bloch equations for nearly adiabatic driving $d\tilde{\rho}_{gg}^{(n)}/dt = -2\Im m[(w_{ge}^{(n)})^*\tilde{\rho}_{ge}^{(n)}] -(\Gamma_{ge}^{(n)}+\Gamma_{eg}^{(n)})\tilde{\rho}_{gg}^{(n)} + \Gamma_{eg}^{(n)}$ and $d\tilde{\rho}_{ge}^{(n)}/dt = iw_{ge}^{(n)}(2\tilde{\rho}_{gg}^{(n)}-1) + i(w_{ee}^{(n)}-w_{gg}^{(n)})\tilde{\rho}_{ge}^{(n)} + i\omega_{01}^{(n)} \tilde{\rho}_{ge}^{(n)} - (\Gamma_{eg}^{(n)}/2+\Gamma_{ge}^{(n)}/2 + 2 \Gamma_{\varphi}^{(n)})\tilde{\rho}_{ge}^{(n)}$, where $\tilde{\rho}_{kl}^{(n)} = \braket{k|\hat{\tilde{\rho}}_S^{(n)}|l} = \braket{k^{(n)}|\hat{\rho}_S|l^{(n)}}$, $\hbar \omega_{01}^{(n)} = E_e^{(n)} - E_g^{(n)}$ such that $E_e^{(n)} = \braket{e|\hat{\tilde{H}}_S^{(n)}|e}$ and $E_g^{(n)} = \braket{g|\hat{\tilde{H}}_S^{(n)}|g}$, $w_{kl}^{(n)} = \braket{k|\hat{w}_n|l}$, $\Gamma_{ge}^{(n)} = |\braket{e^{(n)}|\hat{Y}|g^{(n)}}|^2 S(-\omega_{01}^{(n)})$, $\Gamma_{eg}^{(n)} = |\braket{e^{(n)}|\hat{Y}|g^{(n)}}|^2 S(\omega_{01}^{(n)})$, $\Gamma_{\varphi}^{(n)} = |\braket{g^{(n)}|\hat{Y}|g^{(n)}}|^2 S(0)$, where we denote $S(\omega') = \int_{-\infty}^{\infty} d\tau \mathrm{Tr}_E \{\hat{\rho}_E \hat{X}_I(\tau) \hat{X}_I(0)\} e^{i\omega' \tau}/\hbar^2$~\footnote{The conventional notation for operators in the interaction picture is used such that $\hat{X}_I(t) = e^{i\hat{H}_E t/\hbar} \hat{X} e^{-i\hat{H}_E t/\hbar}$.}. Importantly, we assumed $\braket{g^{(n)}|\hat{Y}|g^{(n)}} = -\braket{e^{(n)}|\hat{Y}|e^{(n)}}$ as well as the feasibility of the secular approximation for the $(n+1)$th order nonsteered evolution to neglect the nonsecular terms originally included in Ref.~\cite{pra82/062112}. The approximation is justified in our analysis as we are interested in the population dynamics rather than small contributions to the coherences. The error in this master equation is of the third order in the system--environment coupling strength and of the first order in $\alpha_{n}$.

The master equation above translates to
\begin{equation}
\begin{split}
\dot{\hat{\tilde{\rho}}}^{(n)}_S &= -\frac{i}{\hbar} \left[\hat{\tilde{H}}^{(n)}_S + \hbar \hat{w}_n,  \hat{\tilde{\rho}}^{(n)}_S \right] \\
 &+  \sum_{i=0}^2 \left( \hat{\tilde{L}}_{(n,i)} \hat{\tilde{\rho}}^{(n)}_S  \hat{\tilde{L}}_{(n,i)}^\dagger - \frac{1}{2}  \left\lbrace \hat{\tilde{L}}_{(n,i)}^\dagger \hat{\tilde{L}}_{(n,i)}   ,\hat{\tilde{\rho}}^{(n)}_S \right\rbrace \right),
\end{split}
\label{eq:MEA}
\end{equation}
where the right-hand side is defined by the Lindblad operators
\begin{equation}
\begin{split}
\hat{\tilde{L}}_{(n,0)} &= \sqrt{\Gamma_{eg}^{(n)}} \ket{g} \bra{e}, \\
\hat{\tilde{L}}_{(n,1)} &= \sqrt{\Gamma_{ge}^{(n)}} \ket{e} \bra{g}, \\
\hat{\tilde{L}}_{(n,2)} &= \sqrt{\Gamma_{\varphi}^{(n)}} (\ket{e} \bra{e}-\ket{g} \bra{g}).
\end{split}
\end{equation}
Notice that the component form is immediately retrieved as $\braket{k|\dot{\hat{\tilde{\rho}}}^{(n)}_S|l} = \braket{k|\partial_t [(\hat{D}_S^{(n)})^{\dagger} \hat{\rho}_S (\hat{D}_S^{(n)})]|l} = \partial_t [\braket{k|(\hat{D}_S^{(n)})^{\dagger} \hat{\rho}_S (\hat{D}_S^{(n)})|l} = \partial_t \tilde{\rho}_{kl}^{(n)}$ since $\ket{k},\ket{l}$ are fixed in time. To obtain a straightforward interpretation of the driven evolution for the Monte Carlo wavefunction method used in our stochastic analysis, we define a transformation back to the original Schr\"odinger picture as $\hat{A}^{B} = (\hat{D}_S^{(n)}) \hat{A} (\hat{D}_S^{(n)})^{\dagger}$ for arbitrary operator $\hat{A}$. The time-derivate of $\hat{\tilde{\rho}}^{(n)}_S$ deconstructs as
\begin{equation}
\begin{split}
\dot{\hat{\tilde{\rho}}}^{(n)}_S &= \partial_t [(\hat{D}_S^{(n)})^{\dagger}] \hat{\rho}_S (\hat{D}_S^{(n)}) \\ &+ (\hat{D}_S^{(n)})^{\dagger} \hat{\rho}_S \partial_t [(\hat{D}_S^{(n)})] \\ &+ ( \hat{D}_S^{(n)})^{\dagger} \dot{\hat{\rho}}_S (\hat{D}_S^{(n)}),
\end{split}
\end{equation}
and hence
\begin{equation}
\begin{split}
(\dot{\hat{\tilde{\rho}}}^{(n)}_S)^B = \dot{\hat{\rho}}_S + [(\hat{D}_S^{(n)}) \partial_t [(\hat{D}_S^{(n)})^{\dagger}],\hat{\rho}_S],
\label{eq:BW}
\end{split}
\end{equation}
where we used the unitarity of $\hat{D}_n$ for every $n$. Back transforming Eq.~(\ref{eq:MEA}) with the help of Eq.~(\ref{eq:BW}) results in Eq.~(\ref{eq:ME}) in the main text. Note that the unitary part in Eq.~(\ref{eq:ME}) is retrieved from $\hat{D}_S^{(n)}[\hat{\tilde{H}}^{(n)}_S + \hbar \hat{w}_n](\hat{D}_S^{(n)})^{\dagger} + \hbar \hat{W}_n  = \hat{H}_S$, where $\hat{W}_n = -i(\hat{D}_S^{(n)}) \partial_t [(\hat{D}_S^{(n)})^{\dagger}]$ stems from Eq.~(\ref{eq:BW}).

\section{Derivation of the integral fluctuation theorem in Eq.~(\ref{eq:RI})} \label{app:IFT}

Let us consider an open quantum system whose dynamics are determined by a Lindblad equation of the form
\begin{equation}
\begin{split}
\dot{\hat{\rho}}_S &= -\frac{i}{\hbar} \left[ \hat{H}_S,  \hat{\rho}_S \right] \\
 &+  \sum_{i=0}^{M} \left( \hat{L}_{i} \hat{\rho}_S  \hat{L}_{i}^\dagger - \frac{1}{2}  \left\lbrace \hat{L}_{i}^\dagger \hat{L}_{i}   ,\hat{\rho}_S \right\rbrace \right),
\end{split}
\label{eq:ME2}
\end{equation}
where the number of dissipative channels is $M+1$ and the Lindblad operator corresponding to the channel $i$ is given by $\hat{L}_{i}(t)=\sqrt{\Gamma_i(t)} \hat{A}_i(t)$. Every operator in this expression is potentially time dependent. To exactly unravel the dynamics, we apply the quantum jump method with the two-measurement approach as described in Sec.~\ref{sec:Qtraj} of the main text such that each event takes place over an infinitesimal time interval $\delta t$. The probability of the system to traverse a single $N$-jump trajectory is given by
\begin{equation}
\begin{split}
& P_{\rightarrow}[E_k^{(1)}(t_{\mathrm{init}}), E_l^{(1)}(t_{\mathrm{final}}), \{\hat{L}_{m_j}\}_{j=1}^N, \{t_j\}_{j=1}^N] \\ &= P[E_k^{(1)}(t_{\mathrm{init}})] \left( \prod_{j=1}^N p_{m_j}(t_j) p^0(t_j,t_{j-1}) \right) \\ &\times p^0(t_{\mathrm{final}}, t_N) P[E_l^{(1)}(t_{\mathrm{final}})],
\label{Eq:Pr1}
\end{split}
\end{equation}
where $E_k^{(1)}(t_{\mathrm{init}})$ and $E_l^{(1)}(t_{\mathrm{final}})$ are the measured eigenenergies in the TMA for this particular realization, $P[E_k^{(1)}(t_{\mathrm{init}})]$ and $P[E_l^{(1)}(t_{\mathrm{final}})]$ are the corresponding probabilities, respectively, $p_{m_j}(t_j)$ is the probability for a jump event to occur along the $m_j$th channel during $[t_j,t_j+\delta t]$, and $p^0(t_{j+1},t_j)$ is the probability of no jumps to occur during the time-interval $[t_j,t_{j+1}]$, that is, between two subsequent jump events. Note that we use the notation $t_0 = t_{\mathrm{init}}$ in the product to account for the no-jump evolution between the initial projective measurement and the first jump event. In the following, we denote the initial and final eigenstates corresponding to the measured energies as $\ket{k^{(1)} (t_{\mathrm{init}})}$ and $\ket{l^{(1)}(t_{\mathrm{final}})}$, respectively. All of the probabilities on the right-hand side of Eq. \eqref{Eq:Pr1} beyond that corresponding to the initial measurement are conditional in the sense that they depend on the earlier traversal history.

Similarly to the main text, the no-jump evolution during the time interval $[t_j,t_{j+1}]$ is given by
\begin{equation}
\hat{U}_{\mathrm{eff}}(t_{j+1},t_{j})= \mathcal{T} \exp \left\{ -\frac{i}{\hbar}\int_{t_{j}}^{t_{j+1}} \hat{H}_{\mathrm{eff}}(t) dt \right\},
\end{equation}
where the non-Hermitian effective Hamiltonian is $\hat{H}_{\mathrm{eff}}(t)=\hat{H}_S(t)-i \hbar \sum_{i} \hat{L}_i^\dagger(t) \hat{L}_i(t)/2$. The state after the no-jump evolution is
\begin{align}
|\psi(t_{j+1})\rangle =  \frac{\hat{U}_{\mathrm{eff}}(t_{j+1},t_j)}{\sqrt{p^0(t_{j+1},t_{j})}} |\phi_{j}(t_j)\rangle,
\label{Eq:Pr3}
\end{align}
where $\ket{\phi_{j}(t_j)}$ is the state after the $j$th jump event has occurred at $t_j$ given by
 \begin{align}
&|\phi_{{j}}(t_j)\rangle= \frac{\hat{L}_{m_j}}{\sqrt{ p_{{j}}(t_j)/\delta t}} |\psi(t_{j})\rangle,
\label{Eq:Pr4}
\end{align}
where the probability of the jump event to take place along the $m_j$th channel is given by
\begin{equation}
\begin{split}
p_{m_j}(t_{j}) &= \delta t \langle \psi(t_{j})| \hat{L}^\dagger_{m_j} \hat{L}_{m_j} |\psi(t_{j})\rangle \\
&=  \frac{\delta t}{p^0(t_{j},t_{j-1})} \langle \phi_{{j-1}}(t_{j-1})|\hat{U}_{\mathrm{eff}}^\dagger(t_{j},t_{j-1}) \\ &\times \hat{L}^\dagger_{m_j} \hat{L}_{m_j} \hat{U}_{\mathrm{eff}}(t_{j},t_{j-1}) |\phi_{{j-1}} (t_{j-1})\rangle.
\end{split}
\label{Eq:pmj}
\end{equation}
In addition, the probability for the no-jump evolution occurring can be written as~\cite{prl68/580, pra45/4879, pra46/4363, OSAQO, rmp70/101}
\begin{align}
p^0(t_{j+1},t_{j}) =  || \hat{U}_{\mathrm{eff}}(t_{j+1},t_j) |\phi_{{j}}(t_j)\rangle ||^2.
\label{Eq:Pr2}
\end{align}
The above notation accounts for the no-jump evolution after the initial measurement and the first jump event by adopting the convention $\ket{\phi_{0}(t_0)} = \ket{k^{(1)}(t_{\mathrm{init}})}$.

The final projective measurement is carried out after a no-jump evolution has occurred during the time interval $[t_N,t_{\mathrm{final}}]$ and yields $E_l^{(1)}(t_{\mathrm{final}})$ with the probability
\begin{align}
P[E_l^{(1)}(t_{\mathrm{final}})]= \frac{| \langle l^{(1)}(t_{\mathrm{final}})| \hat{U}_{\mathrm{eff}}(t_{\mathrm{final}},t_N) |\phi_{{N}}(t_N)\rangle |^2} {p^0(t_{\mathrm{final}},t_N)}.
\label{Eq:Pr5}
\end{align}
Using Eqs.~\eqref{Eq:pmj}--\eqref{Eq:Pr5}, we have the identities
\begin{equation}
\begin{split}
&p_{m_j}(t_j) p^0(t_j,t_{j-1}) \\ &= \delta t  \langle \phi_{{j-1}}(t_{j-1})|\hat{U}_{\mathrm{eff}}^\dagger(t_{j},t_{j-1}) \hat{L}^\dagger_{m_j} \\ &\times \hat{L}_{m_j} \hat{U}_{\mathrm{eff}}(t_{j},t_{j-1}) |\phi_{{j-1}} (t_{j-1})\rangle,
\end{split}
\end{equation}
and
\begin{equation}
\begin{split}
&p^0(t_{\mathrm{final}}, t_N) P[E_l^{(1)}(t_{\mathrm{final}})] \\ &= \big| \langle l^{(1)}(t_{\mathrm{final}})| \hat{U}_{\mathrm{eff}}(t_{\mathrm{final}},t_N) |\phi_{{N}}(t_N)\rangle \big|^2.
\end{split}
\end{equation}
Hence, the traversal probability for the $N$-jump trajectory in Eq.~\eqref{Eq:Pr1} takes the form
\begin{widetext}
\begin{equation}
\begin{split}
& P_{\rightarrow}[E_k^{(1)}(t_{\mathrm{init}}), E_l^{(1)}(t_{\mathrm{final}}), \{\hat{L}_{m_j}\}_{j=1}^N, \{t_j\}_{j=1}^N] \\ &= P[E_k^{(1)}(t_{\mathrm{init}})] \big| \langle l^{(1)}(t_{\mathrm{final}})| \hat{U}_{\mathrm{eff}}(t_{\mathrm{final}},t_N) |\phi_{{N}}(t_N)\rangle \big|^2 \prod_{j=1}^N \delta t  \langle \phi_{{j-1}}(t_{j-1})|\hat{U}_{\mathrm{eff}}^\dagger(t_{j},t_{j-1}) \hat{L}^\dagger_{m_j} \hat{L}_{m_j} \hat{U}_{\mathrm{eff}}(t_{j},t_{j-1}) |\phi_{{j-1}} (t_{j-1})\rangle \\
&= (\delta t)^N P[E_k^{(1)}(t_{\mathrm{init}})] \big| \langle l^{(1)}(t_{\mathrm{final}})| \hat{U}_{\mathrm{eff}}(t_{\mathrm{final}},t_{N}) [ \prod_{j=1}^N \hat{L}_{m_{N+1-j}}\hat{U}_{\mathrm{eff}}(t_{N+1-j},t_{N-j})]|k^{(1)}(t_{\mathrm{init}}) \rangle \big|^2 \\
&=  (\delta t)^N P[E_k^{(1)}(t_{\mathrm{init}})] [\prod_{i=1}^N \Gamma_{m_i}] \big|  \langle l^{(1)}(t_{\mathrm{final}}) | \hat{U}_{\mathrm{eff}}(t_{\mathrm{final}},t_N) [ \prod_{j=1}^N \hat{A}_{m_{N+1-j}}\hat{U}_{\mathrm{eff}}(t_{N+1-j},t_{N-j})]|k^{(1)}(t_{\mathrm{init}}) \rangle \big|^2,
\label{Eq:Pr6}
\end{split}
\end{equation}
\end{widetext}
where after the second equality we used the identity $\ket{\phi_j(t_j)}\bra{\phi_j(t_j)}\hat{L}_{m_j} = \hat{L}_{m_j}$ stemming from Eq.~(\ref{Eq:Pr4}), we denote $\hat{L}_{m_j} = \hat{L}_{m_j}(t_j)$, $\Gamma_{m_j} = \Gamma_{m_j}(t_j)$, and $\hat{A}_{m_j} = \hat{A}_{m_j}(t_j)$, and the ordering of the operator products is defined the same way as in Sec.~\ref{sec:AR}.

In order to formulate the integral fluctuation theorem, we follow the standard approach~\cite{pre73/046129,jsp148/480,PhysRevE.85.031110,NJP15/085028,PhysRevA.88.042111,s10955-014-0991-1,rmp81/1665,PTRSA.369.291} and study the time-reversed counterparts of the trajectories generated by Eq.~(\ref{eq:ME2}). We denote reversed time by $\bar{t}=(t_{\mathrm{init}} + t_{\mathrm{final}})-t$. Moreover, we define the $N$-jump time-reversed trajectory corresponding to the time-forward one presented above by carrying out the first measurement in the TMA at $\bar{t} = t_{\mathrm{init}}$ yielding $\bar{E}_l^{(1)}(t_{\mathrm{init}})$ and the second measurement at $\bar{t} = t_{\mathrm{final}}$ yielding $\bar{E}_k^{(1)}(t_{\mathrm{final}})$, where we denote $\bar{E}_s^{(1)}(\bar{t}) = E_s^{(1)}(t)$, $s\in \{k,l\}$. We denote the corresponding energy eigenstates by $\ket{\bar{s}(\bar{t})} = \ket{s(t)}$, $s\in \{k,l\}$. The time-reversed trajectory is traversed by reversing the system dynamics  and by imposing $N$ quantum jumps occurring at times $ \tau_j = (t_{\mathrm{init}} + t_{\mathrm{final}})-t_{N+1-j}$ due to time-reversed jump operators $\hat{\bar{L}}_{\bar{m}_j}(\tau_j)=\sqrt{\bar{\Gamma}_{\bar{m}_j}(\tau_j)}\hat{A}_{\bar{m}_j}^\dagger(t_{N+1-j})$, where the index $\bar{m}_j$ is related to the indexing of the forward trajectory as  $\bar{m}_j = m_{N+1-j}$~\footnote{Due to this relation, the set of possible values for $\bar{m}_j$  is the same as for $m_j$.} and $\bar{\Gamma}_{\bar{m}_j}$ denotes the transition rate of the time-reversed jump. At this point, we assume that the values of the reversed transition rates can be chosen freely. They will be specified later by requiring Eq. \eqref{Eq:LL} to be satisfied. Note that $\tau_j$ is formulated in such a way that the reversed-time jump events occur in increasing order. The unitary part of time-reversed evolution is governed by $\hat{\bar{H}}_S(\bar{t})=-\hat{H}_S(t)$, where the minus sign reverses the intrinsic system dynamics. Therefore, the time-reversed trajectories correspond to the unraveling of the Lindblad equation
\begin{equation}
\begin{split}
\frac{d {\hat{\bar{\rho}}}_S}{d \bar{t} } &= -\frac{i}{\hbar} \left[ \hat{\bar{H}}_S(\bar{t}),  \hat{\bar{\rho}}_S(\bar{t}) \right] \\
 &+  \sum_{i=0}^{M} \left( \hat{\bar{L}}_{i} (\bar{t}) \hat{\bar{\rho}}_S(\bar{t}) \hat{\bar{L}}_{i}^\dagger(\bar{t}) - \frac{1}{2}  \left\lbrace \hat{\bar{L}}_{i}^\dagger(\bar{t}) \hat{\bar{L}}_{i}(\bar{t}),\hat{\bar{\rho}}_S(\bar{t}) \right\rbrace \right),
\end{split}
\label{eq:ME3}
\end{equation}
where $\hat{\bar{L}}_{i}(\bar{t})=\sqrt{\bar{\Gamma}_{i}(\bar{t})}\hat{A}_{i}^\dagger(t)$. Equation \eqref{eq:ME3} guarantees that the time-reversed trajectories are true quantum trajectories.

Similarly to the time-forward trajectory, the state after the $j$th jump in the time-reversed trajectory is given by
\begin{align}
|\bar{\phi}_{{j}}(\tau_j)\rangle= \frac{\hat{\bar{L}}_{\bar{m}_j}}{\sqrt{ \bar{p}_{{j}}(\tau_j)/\delta t}} |\bar{\psi}(\tau_{j})\rangle,
\label{Eq:rphi}
\end{align}
where the jump probability is
\begin{align}
\bar{p}_{\bar{m}_j}(\tau_j)= \delta t \langle \bar{\psi}(\tau_j)|\hat{\bar{L}}_{\bar{m}_j}^\dagger \hat{\bar{L}}_{\bar{m}_j} |\bar{\psi}(\tau_j)\rangle,
\end{align}
and the state after the no-jump evolution is given by
\begin{equation}
|\bar{\psi}(\tau_{j+1})\rangle =  \frac{\hat{\bar{U}}_{\mathrm{eff}}(\tau_{j+1},\tau_{j})}{\sqrt{\bar{p}^{0}(\tau_{j+1},\tau_{j})}} \ket{\bar{\phi}_j(\tau_j)}.
\end{equation}
Here the time-reversed no-jump evolution is governed by
\begin{equation}
\hat{\bar{U}}_{\mathrm{eff}}(\tau_{j+1},\tau_{j})=\bar{\mathcal{T}} \exp \left\{ -\frac{i}{\hbar}\int_{\tau_{j}}^{\tau_{j+1}} \hat{\bar{H}}_{\mathrm{eff}}(\bar{t}) d\bar{t} \right\},
\end{equation}
where $\hat{\bar{H}}_{\mathrm{eff}}(\bar{t}) = \hat{\bar{H}}_{S}(\bar{t})-i \hbar \sum_{i} \hat{\bar{L}}_{i}^\dagger(\bar{t}) \hat{\bar{L}}_{i}(\bar{t})/2$ and $\bar{\mathcal{T}}$ denotes time ordering in reversed time. Note that this effective Hamiltonian accounts for the reversal of the control protocol and the reversed jump processes. The control protocol is given in real time and, hence, time-reversal is applied for $\hat{H}_S$, whereas the jump operators are defined in reversed time by default. In addition, the no-jump probability is obtained from
\begin{equation}
\bar{p}^{0}(\tau_{j+1},\tau_{j}) =  || \hat{\bar{U}}_{\mathrm{eff}}(\tau_{j+1},\tau_{j}) |\bar{\phi}_{j}(\tau_j)\rangle ||^2.
\end{equation}

The probability to traverse the above-defined time-reversed trajectory is obtained similarly to the time-forward one and takes the form
\begin{equation}
\begin{split}
&\bar{P}_{\rightarrow}[\bar{E}_l^{(1)}(t_{\mathrm{init}}), \bar{E}_k^{(1)}(t_{\mathrm{final}}), \{\hat{\bar{L}}_{\bar{m}_j} \}_{j=1}^N, \{\tau_j\}_{j=1}^N] \\
&= (\delta t)^N \bar{P}[\bar{E}_l^{(1)}(t_{\mathrm{init}})] [\prod_{i=1}^N  \bar{\Gamma}_{\bar{m}_i} ]  \big|  \langle \bar{k}^{(1)}(t_{\mathrm{final}}) |\hat{\bar{U}}_{\mathrm{eff}}(t_{\mathrm{final}},\tau_{N}) \\ & \times[ \prod_{j=1}^N \hat{A}^{\dagger}_{m_j}\hat{\bar{U}}_{\mathrm{eff}}(\tau_{N+1-j},\tau_{N-j})]|\bar{l}^{(1)}({t}_{\mathrm{init}}) \rangle\big|^2,
\end{split}
\label{Eq:Pr8}
\end{equation}
where we denote $\hat{\bar{L}}_{\bar{m}_j} = \hat{\bar{L}}_{\bar{m}_j}(\tau_j)$, $\bar{\Gamma}_{\bar{m}_j} = \bar{\Gamma}_{\bar{m}_j}(\tau_j)$, and $\hat{A}_{m_j} = \hat{A}_{m_j}(t_j)$. If we assume for all $t$ the following condition
\begin{equation}
\sum_{i=0}^M \hat{\bar{L}}_{i}^{\dagger}(\bar{t}) \hat{\bar{L}}_{i}(\bar{t})= \sum_{i=0}^M \hat{L}_{i}^\dagger(t) \hat{L}_{i}(t),
\label{Eq:LL}
\end{equation}
where the summation is over all dissipative channels, the forward and reversed in time no-jump evolution operators can be shown to fulfill
\begin{equation}
\hat{\bar{U}}_{\mathrm{eff}}(\tau_{j+1},\tau_{j}) = \hat{U}_{\mathrm{eff}}^\dagger(t_{N+1-j},t_{N-j}),
\label{Eq:CC}
\end{equation}
where one should be mindful of the transformation $ \tau_j = (t_{\mathrm{init}} + t_{\mathrm{final}})-t_{N+1-j}$ when applying the expression. For example, in the case of a Lindblad equation which contains a decay channel $\hat{A}_k(t)=\hat{A}_i^\dagger(t)$ for each channel $\hat{A}_i$, Eq.~\eqref{Eq:LL} is satisfied by defining the reversed transitions such that $\bar{\Gamma}_{i}(\bar{t})=\Gamma_{k}(t)$. With this assumption, Eqs.~(\ref{Eq:Pr6}) and (\ref{Eq:Pr8}) allow us to write the trajectory-dependent entropy production~\cite{pre73/046129} as
\begin{equation}
\begin{split}
&R[E_k^{(1)}(t_{\mathrm{init}}), E_l^{(1)}(t_{\mathrm{final}}), \{\hat{L}_{m_j}\}_{j=1}^N, \{t_j\}_{j=1}^N] \\ &= \ln \left[ \frac{P_{\rightarrow}(E_k^{(1)}(t_{\mathrm{init}}), E_l^{(1)}(t_{\mathrm{final}}), \{\hat{L}_{m_j}\}_{j=1}^N, \{t_j\}_{j=1}^N)}{\bar{P}_{\rightarrow}(\bar{E}_l^{(1)}(t_{\mathrm{init}}), \bar{E}_k^{(1)}(t_{\mathrm{final}}), \{\hat{\bar{L}}_{\bar{m}_j}\}_{j=1}^N, \{\tau_j\}_{j=1}^N)} \right] \\
&=\ln \left[  \frac{P[E_k^{(1)}(t_{\mathrm{init}})]}{\bar{P}[\bar{E}_l^{(1)}(t_{\mathrm{init}})]} \prod_{j=1}^{N} \frac{\Gamma_{m_j}(t_j)}{\bar{\Gamma}_{\bar{m}_j}(\tau_j)}\right] \\
&=\ln \left[  \frac{P[E_k^{(1)}(t_{\mathrm{init}})]}{\bar{P}[\bar{E}_l^{(1)}(t_{\mathrm{init}})]} \prod_{j=1}^{N} \frac{\Gamma_{m_j}(t_j)}{\bar{\Gamma}_{m_j}(\bar{t}_j)}\right],
\end{split}
\label{Eq:R}
\end{equation}
where the product counts over all jump events in the trajectories and in the last line the product was re-organized using the relations $\bar{m}_j = m_{N+1-j}$ and $\bar{t}_j = (t_{\mathrm{init}} + t_{\mathrm{final}}) - t_j$. This allows a more convenient expression in terms of comparing individual jump processes between the trajectories.

Following the conventional approach to fluctuation relations in, e.g., Refs.~\cite{Seifert2005,rmp81/1665}, we define the probability distribution of the trajectory-dependent entropy production in Eq.~(\ref{Eq:R}) for forward-time evolution as
\begin{equation}
\begin{split}
& d(R) = \sum_{k,l} \sum_{N=0}^{\infty}  \sum_{\{\hat{L}_{m_j}\}_{j=1}^N} \frac{1}{N!} \prod_{i=1}^N \int_{t_{\mathrm{init}}}^{t_{\mathrm{final}}}dt_i \\ &\times \frac{1}{(\delta t)^N} P_{\rightarrow}[E_k^{(1)}(t_{\mathrm{init}}), E_l^{(1)}(t_{\mathrm{final}}), \{\hat{L}_{m_j}\}_{j=1}^N, \{t_j\}_{j=1}^N] \\ &\times \delta \{R-R[E_k^{(1)}(t_{\mathrm{init}}), E_l^{(1)}(t_{\mathrm{final}}), \{\hat{L}_{m_j}\}_{j=1}^N, \{t_j\}_{j=1}^N]\},
\label{eq:pR}
\end{split}
\end{equation}
where the first summation is over all possible initial and final measurement results, the second summation is over all possible numbers of jumps a single trajectory can undergo, and the third summation is over all possible sets of jump operators with $N$ jumps. The integrations account for all possible jump times during an $N$-jump trajectory. The corresponding expression for the reversed-time evolution is
\begin{equation}
\begin{split}
& \bar{d}(R) = \sum_{k,l} \sum_{N=0}^{\infty}  \sum_{\{\hat{\bar{L}}_{\bar{m}_j}\}_{j=1}^N} \frac{1}{N!} \prod_{i=1}^N \int_{t_{\mathrm{init}}}^{t_{\mathrm{final}}}d\tau_i \\ &\times \frac{1}{(\delta t)^N} \bar{P}_{\rightarrow}[\bar{E}_l^{(1)}(t_{\mathrm{init}}), \bar{E}_k^{(1)}(t_{\mathrm{final}}), \{\hat{\bar{L}}_{\bar{m}_j}\}_{j=1}^N, \{\tau_j\}_{j=1}^N] \\ &\times \delta \{R+R[E_k^{(1)}(t_{\mathrm{init}}), E_l^{(1)}(t_{\mathrm{final}}), \{\hat{L}_{m_j}\}_{j=1}^N, \{t_j\}_{j=1}^N]\},
\label{eq:pRrev}
\end{split}
\end{equation}
where we have used the fact that the entropy productions in the reversed and forward trajectories have the same magnitude but different sign stemming from Eq.~(\ref{Eq:R}). As a direct consequence of Eqs.~(\ref{Eq:R})--(\ref{eq:pRrev}), the following detailed fluctuation theorem holds
\begin{align}
\frac{d(R)}{ \bar{d}(-R)}=e^R. \label{Eq:DFT}
\end{align}

The only assumptions needed for Eq. \eqref{Eq:DFT} are that the condition in Eq.~(\ref{Eq:LL}) is fulfilled and that $\{ \ket{k^{(1)}_{\mathrm{init/final}} (t_{\mathrm{init/final}}) } \}$ forms a basis for the system, that is, that the initial and final measurements are able to probe the corresponding full Hilbert space. Equation~\eqref{Eq:DFT} implies that the integral fluctuation theorem holds such that
\begin{align}
\braket{ e^{-R} } =1,
\label{Eq:JR}
\end{align}
where $\braket{\dots}$ denotes an ensemble average over the $R$-distribution given in Eq.~(\ref{eq:pR}). Inserting the definition of $R$ in Eq.~(\ref{Eq:R}) into Eq.~(\ref{Eq:JR}) yields the final form of the integral fluctuation theorem as
\begin{equation}
\Braket{  \frac{\bar{P}[\bar{E}_l^{(1)}(t_{\mathrm{init}})]}{P[E_k^{(1)}(t_{\mathrm{init}})]} \prod_{j=1}^{N} \frac{\bar{\Gamma}_{m_j}(\bar{t}_j)}{\Gamma_{m_j}(t_j)} } = 1.
\end{equation}

Equation~(\ref{eq:ME}) is in the Lindlabd form and using Eq.~(\ref{eq:L}) it can be shown that the condition in Eq.~(\ref{Eq:LL}) is fulfilled for the sets $\{ \hat{L}_{(n,i)}(t) \}_{i=0}^2$ and $\{ \hat{\bar{L}}_{(n,i)}(\bar{t}) \}_{i=0}^2$ by asserting $\bar{\Gamma}_{(n,0)}(\bar{t})=\Gamma_{(n,1)}(t)$, $\bar{\Gamma}_{(n,1)}(\bar{t})=\Gamma_{(n,0)}(t)$, and $\bar{\Gamma}_{(n,2)}(\bar{t})=\Gamma_{(n,2)}(t)$. Hence, the integral fluctuation theorem takes the form of Eq.~(\ref{eq:RI}) in the main text. Note especially that in this case the stochastic dynamics is generated by $n$th-order master equation and correspondingly the transition rates and the ensemble average depend on $n$.

\section{Probability distribution of work in the $n$th-order approximation} \label{app:pnw}

Using the notation introduced in Appendix~\ref{app:IFT}, the probability distribution of work using the $n$th-order dynamics can generally be written as
\begin{equation}
\begin{split}
& d_n(W^{(n)}) = \sum_{k,l} \sum_{N=0}^{\infty}  \sum_{\{\hat{L}_{m_j}\}_{j=1}^N} \frac{1}{N!} \prod_{i=1}^N \int_{t_{\mathrm{init}}}^{t_{\mathrm{final}}}dt_i \\ &\times \frac{1}{(\delta t)^N} P^{(n)}_{\rightarrow}[E_k^{(1)}(t_{\mathrm{init}}), E_l^{(1)}(t_{\mathrm{final}}), \{\hat{L}_{m_j}\}_{j=1}^N, \{t_j\}_{j=1}^N] \\ &\times \delta \{W^{(n)}-W^{(n)}[E_k^{(1)}(t_{\mathrm{init}}), E_l^{(1)}(t_{\mathrm{final}}), \{\hat{L}_{m_j}\}_{j=1}^N, \{t_j\}_{j=1}^N]\},
\label{eq:pnW}
\end{split}
\end{equation}
where the trajectory-dependent work $W^{(n)}$ is given in Eq.~(\ref{eq:Wn}) and $P^{(n)}_{\rightarrow}$ is given in Eq.~(\ref{Eq:Pr6}) such that the available dissipative channels correspond to the set of Lindblad operators $\{ \hat{L}_{(n,i)}(t) \}_{i=0}^2$ in Eq.~(\ref{eq:L}). Note especially that both $P^{(n)}_{\rightarrow}$ and $W^{(n)}$ depend on the number of transformations through the jump operators.
\vspace{10 cm}
\bibliography{localbib}

\end{document}